\documentclass[preprint,11pt]{elsarticle}
\usepackage{lipsum}
\makeatletter
\def\ps@pprintTitle{
 \let\@oddhead\@empty
 \let\@evenhead\@empty
 \def\@oddfoot{}
 \let\@evenfoot\@oddfoot}
 \renewcommand{\MaketitleBox}{
  \resetTitleCounters
  \def\baselinestretch{1}
  \begin{center}
    \def\baselinestretch{1}
    \Large \@title \par
    \vskip 18pt
    \normalsize\elsauthors \par
    \vskip 10pt
    \footnotesize \itshape \elsaddress \par
  \end{center}
  \vskip 12pt
}

\usepackage{amssymb}
\usepackage{amsthm}
\usepackage{lineno}
\usepackage{amsfonts}
\usepackage{amsmath}
\usepackage{graphicx}
\usepackage{cases}
\usepackage{epstopdf}
\usepackage{caption}
\usepackage{algorithmic}
\biboptions{sort&compress}
\ifpdf
  \DeclareGraphicsExtensions{.eps,.pdf,.png,.jpg}
\else
  \DeclareGraphicsExtensions{.eps}
\fi

\usepackage{amssymb}
\usepackage{mathtools}

\usepackage[mathscr]{eucal}
\usepackage{color}
\usepackage{geometry}
\usepackage{tikz}
\usetikzlibrary{shapes,arrows}
\usetikzlibrary{decorations.pathreplacing,angles,quotes}

\usepackage{todonotes}
\usepackage{dsfont}
\usepackage{cleveref}
\usepackage{verbatim}
\usepackage{graphicx}
\usepackage{subfigure}
\usepackage{subcaption}
\usepackage{tikz}
\usepackage{pgfplots}
\usepgfplotslibrary{groupplots}

\newcommand{\mb}{\mathbb}

\newcommand{\mc}{\mathcal}

\newcommand{\mcD}[1]{{\mathcal{D}#1}}

\newtheorem{theorem}{Theorem}[section]

\numberwithin{equation}{section}
\newtheorem{definition}{Definition}[section]

\newcommand{\rulex}{\hfill\rule{1mm}{3mm}}

\begin{document}
\begin{frontmatter}

\title{Large Banks and Systemic Risk: Insights from a Mean-Field Game Model
\footnote{Dena Firoozi would like to acknowledge the support of the Natural Sciences and Engineering Research Council of Canada
(NSERC), grant RGPIN-2022-05337. 
}
\footnote{David Benatia would like to acknowledge the support of the Social Sciences and Humanities Research Council of Canada, grant 430-2022-00544.}}
\author[1]{Yuanyuan Chang}
\author[1]{Dena Firoozi}
\author[2]{David Benatia}
\address[1]{Department of Decision Sciences, HEC Montréal, Montréal, QC, Canada\\ (email: yuanyuan.chang@hec.ca, dena.firoozi@hec.ca)}
\address[2]{Department of Applied Economics, HEC Montréal, HEC Montréal, Montréal, QC, Canada\\ (email: david.benatia@hec.ca)}

\begin{abstract}
This paper presents a dynamic game framework to analyze the role of large banks in interbank markets. By extending existing models, we incorporate a large bank as a dynamic decision-maker interacting with multiple small banks. Using the mean-field game methodology and convex analysis, best-response trading strategies are derived, leading to an approximate equilibrium for the interbank market. We investigate the influence of the large bank on the market stability by examining individual default probabilities and systemic risk, through the use of Monte Carlo simulations. Our findings reveal that, when the size of the major bank is not excessively large, it can positively contribute to market stability. However, there is also the potential for negative spillover effects in the event of default, leading to an increase in systemic risk. The magnitude of this impact is further influenced by the size and trading rate of the major bank. Overall, this study provides valuable insights into the management of systemic risk in interbank markets.
\end{abstract}

\begin{keyword}
Interbank Market, Large Banks, Mathematical Finance, Mean-Field Games, Small Banks, Systemic Risk.
\end{keyword}

\end{frontmatter}

\section{Introduction}
\subsection{Background and Problem Description}
The interbank market serves as a key mechanism for optimizing the usage of monetary reserves and minimizing the amount of capital held in low-return liquid assets by allowing banks to borrow from one another or the central bank to address short-term funding shortages \cite{ Freixas2000SystemicRiskInterbank}. This market also provides a means of risk-sharing between banks, making the individual institutions more resilient to negative shocks \cite{Acemoglu2015SystemicRiskStability}. However, the interbank market also has the potential to create a contagion channel through which the financial distress of one bank can spread to others and potentially lead to a financial crisis \cite{Hautsch2015FinancialNetworkSystemic}. The Canadian overnight market, for example, allows financial institutions to engage in short-term borrowing and lending of funds on an overnight basis, where the interest rate is called the overnight rate. The Bank of Canada sets a target for the overnight rate and  establishes an operating band through its monetary policy operations. While the specific width of the operating band can vary over time, it currently spans a range of 25 basis points (0.25\%). The upper bound of the operating band is the rate at which the Bank of Canada stands ready to lend money to financial institutions, while the lower bound is the rate at which the central bank absorbs excess funds from the system. The overnight rate is determined by the supply and demand dynamics in the interbank market. Financial institutions with excess funds may lend to earn interest on their surplus liquidity, whereas banks facing temporary shortages of funds may borrow to meet their immediate obligations. 

In the aftermath of the 2008 financial crisis, regulatory bodies such as the Financial Stability Board recognized the existence of \emph{Systemically Important Banks} whose stability has a significant impact on the overall economy \citep{Tarashev2009SystemicImportanceFinancial}. To identify such banks, \citep{cont_moussa_santos_2013} suggests a measure based on the Contagion Index of banks. Furthermore, \citep{Laeven2016Banksizecapital} shows that the failure of large banks can lead to disproportionate consequences for the entire financial system. Some studies suggest that larger, well-capitalized banks are less likely to default \citep{Steinbacher2015Howbankscapital}, while others find that the contribution to systemic risk is greater for larger institutions \citep{Laeven2016Banksizecapital}. There is also the potential for large banks to take on more risk if they are perceived as ``too big to fail" \citep{Altunbas2017Realizedbankrisk}. Consequently, effectively managing the risk of financial crises necessitates a deep understanding of the contributions of these major banks to systemic risk. The importance of this understanding has once again been highlighted by recent banking crises in the United States and Switzerland.

To the best of our knowledge, the existing literature has primarily examined the role of a major bank through empirical methods and static models. In contrast to these approaches, our paper presents a distinct perspective by modeling a large bank as a dynamic decision-maker within the interbank market, interacting with multiple small banks. Building upon the work of \cite{CarmonaSysRisk2015}, where the interbank model only includes small banks, we extend the model to incorporate the dynamic interactions involving a major bank. To capture the dynamic behavior of the major bank, we employ the mean-field game methodology, which has recently been developed to accommodate agents with a significant impact on the overall system in a dynamic game setup \cite{Huang2010LargePopulationLQGGames, Nourian2013epsilonNashMean, Carmona2016probabilisticapproachmean,Firoozi2020ConvexanalysisLQG} 

In our model, the major bank engages in borrowing or lending activities with small banks and exercises control over its transaction rate with the central bank to maintain a target level of log-monetary reserves, which could be related to ``reserve requirements'' established by the financial regulator, while navigating the market at minimum cost. To achieve the same objective, a generic small bank involves with borrowing or lending activities with other small banks as well as the major bank and controls its transaction rate with the central bank. We use the convex analysis method developed in \cite{Firoozi2020ConvexanalysisLQG} to derive the equilibrium trading strategies in the interbank market. These strategies lead to an $\epsilon$-Nash equilibrium for the market with a large agent and a finite number of small banks.

Subsequently, we examine the influence of the large bank on the stability of the interbank market under equilibrium conditions. Specifically, we investigate the individual default probability of a small bank and the risk of a systemic event in the presence of a large bank. Following \cite{CarmonaSysRisk2015}, the systemic risk is measured by the probability that the average log-monetary reserve of the whole market, which we refer to as the market state, falls below a predetermined default threshold.

To account for the presence of a large bank, we redefine the market state as a linear combination of the major bank's log-monetary reserves and the average log-monetary reserves of the small banks. The weights assigned to this combination reflect the relative influence of the large bank and the collective influence of the small banks operating in the market. We investigate the contribution of the large bank to both individual and systemic default probabilities by considering three cases: (i) absence of a large bank in the interbank market, (ii) presence of a large bank without default, and (iii) presence of a large bank with default. Although calculating default probabilities and analyzing individual interbank trading data present mathematical challenges and data limitations, we utilize Monte Carlo simulations to compute these probabilities under various scenarios.

Our results provide important insights into the role played by large banks in the stability of the financial system. Our findings indicate that bank size matters, with large banks contributing positively to system stability as long as they are not too big. Large banks provide stability through their ability to allow smaller banks to better coordinate, but they also generate significant negative spillovers in the event of a default, which can dramatically increase systemic risk and offset their positive effect on the system. Additionally, we demonstrate that large banks amplify the dual effects of the interbank market: providing both stability and an increased risk of rare systemic events as banks become more reliant on it. Our results further indicate that these effects are not only exacerbated by the presence of a large bank, but also by its size and its speed of interbank trading.

In summary, this paper makes the following contributions:
\begin{itemize}
    \item To the best of our knowledge, it is the first study to explore the role of a large bank within an interbank market through the lens of a dynamic game framework. By adopting this approach, it offers new insights into the dynamics and implications of major banks in the interbank market.
    \item It introduces the notion of market state in the presence of a major bank, which captures the relative influence of the large bank and the collective impact of small banks operating in the market.
    \item It adapts the variational approach to mean-field games (MFGs) with a major agent while considering the market state as a shared signal among minor agents. This adaptation enables a more intuitive interpretation of the results with respect to individual and systemic default risks. 
    \item Through the use of Monte Carlo simulations, it provides valuable insights into the impact of large banks on the stability of the financial system despite the mathematical challenges in the analytical characterization of systemic risk and  limitations in available interbank trading data.
\end{itemize}

\subsection{Literature Review}
MFG theory has been developed in the early 21st century to model the interactions between a large number of agents \citep{Lasry2007Meanfieldgames,Huang2006Largepopulationstochastic,Huang2007LargePopulationCostCoupledLQG, Carmona2018Probabilistictheorymean}. In such games each agent is not only impacted by its own behavior but also by the mass behaviour of all other agents. 
 MFG theory establishes the existence of approximate Nash equilibria in such games and can be used to obtain the corresponding optimal strategies for each agent in the  system. 
Using various approaches, MFG theory has been extended to model the role played by a major (or influential) agent \citep{Huang2010LargePopulationLQGGames,Nourian2013epsilonNashMean,Carmona2016probabilisticapproachmean,firoozi_epsilon-nash_2021,LASRY2018886,Bensoussan2017,MOON2018200,LiuFirooziBreton2023_arXiv,ThesisDena2019,firoozi2022class} or multiple major agents \citep{FirooziCainesCDC2019}. Furthermore, some literature focuses on the equivalency of solutions to MFG systems with major and minor agents obtained via different approaches \citep{Firoozi2020ConvexanalysisLQG, HuangCIS2020, firoozi_lqg_2022}.

The MFG methodology has been applied to various problems in a wide range of applications, particularly in the context of financial markets, including equilibrium pricing \citep{Firoozi2022MAFI,gomes_mean-field_2018}, optimal investment and execution problems in markets \citep{casgrain_meanfield_2020, FirooziISDG2017,FuUlrich2021,Lacker_Zariphopoulou_Zariphopoulou2019}, compliance market design \citep{firoozi_principal_2022}, and cryptocurrency markets \citep{li2023mean}, just to name a few.
 More related to this paper, several studies of systemic risk use the MFG methodology. The first interbank model using MFGs was proposed in \citep{CarmonaSysRisk2015} with a finite number of small banks borrowing or lending to each other and to the central bank. The log-monetary reserves of the banks are modeled as a system of Ornstein–Uhlenbeck controlled diffusion processes coupled in the drift with the average log-monetary reserve of all banks and subject to correlated noise processes. The paper concludes that interbank transactions improve the stability of the interbank market. \citep{Fouque2013StabilityModelInterbank} uses a set of interacting Feller diffusion processes to model the monetary reserves of banks and quantify the relationship between the lending preference of a bank and its bankruptcy. They conclude that the growth rate and lending preferences are important for understanding the systemic risk in interbank lending. Furthermore, \citep{Fouque2013SystemicRiskIllustrated} shows that interbank borrowing and lending activities increase both the stability and the likelihood of a systemic event. These results are consistent with those from the numerical experiments of \citep{Garnier2013LargeDeviationsMean}. \citep {Sun2018SystemicRiskInterbank} uses the Cox–Ingersoll–Ross process to model the evolution of the log-monetary reserves while \citep{Huang2017RobustStochasticGames} extends the model by incorporating model ambiguity. Furthermore, \citep{Bo2015SystemicRiskInterbanking,sun2022mean,RenDena2024risk,Sophia-thesis} incorporate heterogeneity among banks in terms of model parameters. The work \citep{Bo2015SystemicRiskInterbanking} uses coupled jump diffusion processes to model interaction among banks and provides a useful analytical tool to assess systemic risk. \citep{sun2022mean} verifies the existence of equilibria in a two-group and a multi-group heterogeneous interbank markets. 
Moreover, \citep{RenDena2024risk} and \citep{Sophia-thesis} incorporate risk aversion into interbank models through exponential cost functionals. \citep{RenDena2024risk} uses the Fokker-Planck equation to formulate the probabilities of individual default and systematic risk in response to a common shock, while \citep{Sophia-thesis} conducts an analysis using Monte Carlo simulations in a heterogeneous interbank market that includes a major bank.

Another line of research, related to our work, models banking system as a network and studies contagion and systemic risk based on the network structure (see, for example, \cite{RS-Cont2010, cont_moussa_santos_2013, SR-Network-Hu2012,Contagion-Glasserman2016, ChenLiuYao2016,SR-MS2017,SysRisk-QF-2017-1, interbank-QF-2015, SysRisk-QF-2017-2}). \citep{ContagionCACCIOLI2012} investigates probability of contagion conditional on the failure of the most connected and the biggest banks and shows a targeted policy aimed at reinforcing the stability of the biggest banks improves the stability of the system in certain regimes. \citep{AminiMincaSulem2015,AminiMinca2016} study interbank contagion in financial networks under partial information. \citep{amini2015systemic,Amini-CentralNode2020} consider a financial network with a central clearing counterparty and shows that central counterparty clearing can reduce systemic risk.

We note that the terms ``trading strategy" and ``transaction strategy", ``large bank" and ``major bank", ``small bank" and ``minor bank" are used interchangeably throughout this paper. 

The remainder of this paper is organized as follows. \cref{section:Model Description} introduces the model. \cref{section:MFG formulation} presents the mean-field game framework. \cref{section:BestRespnseTranscation} details the derivation of optimal strategies and the equilibrium. \cref{section:Individual Default and Systemic Risk in Interbank Transactions} introduces definitions of default probabilities and systemic risk. \cref{section:Numerical Experiments} presents numerical experiments about the role played by the large bank. \cref{section:concluding remarks} concludes the paper.

\section{Interbank Market Model}\label{section:Model Description}
In this section, we develop a model of interbank transactions which includes a major bank, a large number of small banks, and the central bank. Each individual private bank is not only assumed to trade with other private banks but can also borrow from or to lend to the central bank.\footnote{In this work, ``lending to the central bank" refers to an individual bank buying Treasury Bonds or acting as a Repo buyer, which the central bank uses as policy instruments to decrease market liquidity.} Private banks use the interbank transaction system and the central bank for borrowing to manage liquidity gaps, or for lending to optimize the return on their available liquidity. In general, they optimize the use of their deposits by keeping a relatively small amount of monetary reserves compared to their capitalization. 

\subsection{Major Bank}
In our setting, we refer to the ``major bank'' as a relatively large influential bank in terms of its market share, which is to be differentiated from a small or minor bank. The major bank is said to be influential because its behavior directly affect the decisions of minor banks. 

We denote the major bank by $\mathcal{A}^0$ and its logarithm of monetary reserves (log-monetary reserves) at time $t$ by $x^0_t$. We consider that if small banks behave competitively, then the major bank has profitable borrowing (lending) opportunities from (to) minor banks if it has lower (higher) log-reserves than the average minor bank. Therefore, the major bank is assumed to borrow from or lend to other banks whenever its log-monetary reserve is, respectively, lower or higher than the average log-monetary reserve across small banks.The log-monetary reserve of the major bank hence satisfies
\begin{gather}
    d x_t^0=a_0\left(x_t^{(N)}-x_t^0\right)d t+u_t^0d t+\sigma_0 d W_t^0, \label{MajorDynamicsOriginal}
\end{gather}
where $x_t^{(N)}=\frac{1}{N}\sum_{i=1}^N x_t^i$. In the above stochastic differential equation (SDE), $x_t^{(N)}$ 
represents the average log-monetary reserves of the large population of minor banks. $u_t^0$ models the borrowing and lending activities of the major bank with the central bank. The instantaneous volume (or rate) of transactions between the major bank and its smaller counterparts is represented by $a_0\big( x_t^{(N)}-x_t^0\big)$. As $a_0$ increases, the major bank trades more with other banks and reverts more quickly to the mean log-monetary reserves of minor banks. The parameter $a_0$ serves as a measure of the degree of market friction and provides an indication of the reliance on the interbank market. This aspect will be further discussed in the subsequent sections. 
The parameter $\sigma_0$ represents the volatility of its log-monetary reserve generated by deposits and withdrawals of retail customers, which we model as the Brownian motion $W^0_t$.

The operational objective of the major bank is  to control its rate (the amount lent or borrowed per unit of time) of borrowing and lending with the central bank, denoted $u_t^0$, and to optimize its deposits by keeping its log-monetary reserve $x_t^0$ close to the average log-monetary reserve of the minor banks $x_t^{(N)}$. Mathematically, the objective of the major bank is to minimize the cost functional
\begin{equation}
\label{MajorCostOriginal}
        J^{[N]}_0(u^0,u^{-0}) = \mathbb{E} \Big[ \int_0^T \Big \{\frac{1}{2} \big(u_t^0\big)^2-q_0 u_t^0\big( x_t^{(N)}- x_t^0\big)+\frac{\epsilon_0}{2}\big(x_t^{(N)}-x_t^0\big)^2 \Big \}d t+\frac{c_0}{2}\big( x_T^{(N)}- x_T^0\big)^2 \Big],
\end{equation}
where $u_t^0$ is the major bank's control variable and $x_t^0$ is its state variable. The optimal strategy chosen by the major bank is represented by $u^0$, and $u^{-0}$ is the collection of the optimal controls of all other banks besides the major bank $u^{-0} = (u^1, \dots, u^N)$. From the cost functional above, the parameter $q_0$ quantifies the incentive to participate in borrowing and lending activity and a higher $q_0$ is akin to the regulator having low fees. $\epsilon_0$ measures the penalization on the major bank when its log-monetary reserve deviates from the average log-monetary reserves of minor banks during the considered period. The parameter $c_0$ penalizes the major bank if there exists a difference between its log-monetary reserves and the average log-monetary reserves of minor banks at the terminal date.

This cost functional consists of 4 terms. The first term $\frac{1}{2} \big(u_t^0\big)^2$ represents a soft constraint on the instantaneous transaction rate with the large bank. The second term $q_0 u_t^0\big( x_t^{(N)}- x_t^0\big)$ models the incentive to trade with the central bank given the current market condition. More specifically, it captures the idea that the interest rate paid for borrowing reserves from the central bank would be smaller if other banks are on average searching for lending activities because demand would be smaller. Reversely, lending to the central bank is more profitable for the major bank if $x_t^0>x_t^{(N)}$ because the supply is smaller. The third term $\frac{\epsilon_0}{2}\big(x_t^{(N)}-x_t^0\big)^2$ is the instantaneous opportunity cost of having too large or too small reserves compared to the other banks. It measures the amount of unrealized profitable transactions, and depends on a parameter $\varepsilon_0\geq0$. Similarly, the last term represents the opportunity cost of holding sub-optimal reserves at the end of time. It is a terminal condition which penalizes deviations from the mean at the final period $T$ using a parameter $c_0\geq0$. We note that one could also consider a model for the major bank aiming to keep its liquidity close to a linear combination of $x^{(N)}_t$ and $x^0_t$. Details on such a model can be found in \cite{Sophia-thesis}.

\subsection{Minor Banks}
In our model, we assume there is a large number $N$ of minor banks in the market. Each minor bank represents a small bank that has a negligible impact on the financial system as the number $N$ grows. We assume that all minor agents are homogeneous, i.e. are statistically identical. We denote a minor bank by $\mathcal{A}^i,i\in \{1,...,N\}, N< \infty$ and its log-monetary reserve at time $t$ by $x^i_t$. The log-monetary reserve $x^i_t$  of minor bank $\mathcal{A}^i$ is assumed to satisfy the SDE
\begin{gather}
    d x_t^i=a\left( \big(F x_t^{(N)}+G x_t^0 \big)-x_t^i\right)dt+u_t^id t+\sigma d W_t^i, \label{MinorDynamicsOriginal}
\end{gather}
where the parameters $F$ and $G$ denote, respectively, the relative size  of the mass of minor banks and of the major bank in the market.

The dynamics of the log-monetary reserve of minor banks is directly influenced by the major bank's state $x_t^0$. A minor bank $\mathcal{A}^i$ optimizes its trading opportunities by keeping an amount of liquidity as close as possible to the market state $\big(F x_t^{(N)}+G x_t^0\big)$. This market state is modeled as a linear combination of the average log-monetary reserve of all minor banks and the log-monetary reserve of the major bank. We assume these relative weights to be common knowledge to all participants and provided by the central bank. 

The term $a\Big(\big(F x_t^{(N)}+G x_t^0 \big)-x_t^i\Big)$ models the volume of transactions of the minor bank with the major bank and other minor banks. The parameter $a$ represents the rate at which a minor bank mean-reverts to the market state through interbank transactions. A higher value of $a$ signifies a greater dependence on the interbank market. $u_t^i$ is the control action of minor bank $i$ and represents the borrowing and lending activities of the minor bank with the central bank. Finally, the parameter $\sigma$ represents the volatility of its log-monetary reserve arising from the activities of their retail customers modeled by the Brownian motion $W^i_t$ at each point $t$ in time. All Brownian motions in this model $W^0_t$ and $W^i_t$ are independent of each other.

Each minor bank chooses the optimal strategy which  minimizes its cost functional defined as
\begin{equation}\label{MinorCostOriginal}
\begin{split}
    J^{[N]}_i(u^i,u^{-i}) = \mathbb{E}\Big[ \int_0^T \Big \{\frac{1}{2} \big(u_t^i\big)^2-q u_t^i\big(F x_t^{(N)}+G x_t^0-x_t^i\big)+\frac{\epsilon}{2}\big(F x_t^{(N)}+G x_t^0-x_t^i\big)^2 \Big \}d t\\
+\frac{c}{2}\big(F x_T^{(N)}+G x_T^0-x_T^i\big)^2 \Big],
\end{split}
\end{equation}
where the strategy chosen by a representative minor bank-$i$ is represented by $u^i$, and the collection of strategies chosen by all other banks is represented by $u^{-i}=(u^0, u^1,\dots,u^{i-1},u^{i+1},\dots,u^{N})$. The cost functional \eqref{MinorCostOriginal} is similar to that of the major bank except for the parameter values and the target level of log-monetary reserves.  In addition, we assume the minor bank's cost functional to be convex by imposing $q^2 \le \epsilon$.

\subsection{Market Clearing Condition}\label{section:Market Clearing Condition}
In equilibrium, the sum of all trades must be equal to zero. In each individual transaction, one bank acts as the lender whereas another acts as the borrower.  The market clearing condition hence states that the total volume of log-monetary reserve transactions should be zero for all $t \in [0,T]$, that is
\begin{equation}\label{ClearingCondition}
  \frac{a}{N} \sum^N_{i=1} \left( \big(F x_t^{(N)}+G x_t^0 \big)-x_t^i\right) +a_0 \left(x_t^{(N)}- x_t^0 \right)=0.
\end{equation}
In the above equation the terms $\frac{a}{N} \sum^N_{i=1} \Big( \big(F x_t^{(N)}+G x_t^0 \big)-x_t^i\Big)$ and $a_0 \big(x_t^{(N)}- x_t^0 \big)$ represent, respectively, the average transactions of minor banks and the transactions of the major bank per unit time with other banks in the market. Rearranging terms from \eqref{ClearingCondition} yields
\begin{equation}
    \big(a F  -a + a_0  \big) x_t^{(N)} + \big(a G - a_0 \big)  x_t^0 =0.
\end{equation}
Note that for the above condition to be satisfied for all $t \in [0,T]$ and for every value of $x_t^{(N)}$ and $x_t^0$, it must be that
\begin{align}
    a_0  &= a - a F, 
    \label{ClearingCondition1}\\
    a_0 &= a G. 
    \label{ClearingCondition2}
\end{align}
Combining these two equations gives $F + G = 1$ 
which supports our interpretation of the parameters $F$ and $G$ as the relative size of the major bank and the mass of minor banks in the market. 

Furthermore, \eqref{ClearingCondition2} characterizes the relationship between the mean reversion rate of the major bank $a_0$ and that of the minor banks $a$ as a function of the relative market size. Note that  the major bank always has a lower mean-reversion rate than minor banks. This occurs in equilibrium because the trade flow of a minor bank is divided into the trades with other minor banks and with the major bank. In equilibrium, the major bank is assigned the share $aG{x}^0_t$ of trades which corresponds to its trade flow $a_0{x}^0_t$ as shown by \eqref{ClearingCondition2}.

We interpret $a_0$ and $a$ as inverse measures of market frictions. A smaller value of $a_0$ means larger market friction for the major bank, which implies a smaller trading rate everything else being equal. The fact that $a_0$ is smaller than $a$ imply that the major bank is subject to more market frictions. It might seem counterintuitive at first since large banks have comparative advantages due to their larger capitalizations. However,  \citet{Bucher2014FrictionsInterbankMarket} and \citet{ArceLargeCentralBank} argue that interbank frictions mainly exist in the form of transaction costs: a bank must find an appropriate counterparty that satisfies two conditions: (1) matching the liquidity requirements and (2) willing to make an agreement. Unlike smaller banks, the major bank has to split large amounts of liquidity into small trades with different (smaller) counterparties, which naturally increases transaction costs. 

\subsection{Equilibrium} 
Each bank in the interbank market makes borrowing and lending decisions based on the information about the states of other banks. Our objective is to identify the optimal borrowing and lending strategies for individual banks in equilibrium. Since each bank interacts with all others and selects an optimal strategy for trading with the central bank in response to these interactions, the concept of Nash equilibria becomes relevant. A Nash equilibrium is achieved when no individual bank can gain an additional benefit by unilaterally changing its strategy, meaning that a bank has no incentive to deviate from a Nash strategy while all other banks are following it. Next, we provide the mathematical definition of Nash equilibrium.

Consider a non-cooperative game with $N$ agents. Each agent-$i$, where $i \in \{1, \dots, N\}$, selects a strategy denoted by $u^i$ from the set of admissible strategies $\mathcal{U}^i$ to minimize its cost functional $J_i(u^i,u^{-i})$. 

\begin{definition}[Nash Equilibrium] An $N$-tuple of strategies $(u^1,\dots,u^{N}) \in \mathcal{U}^{1}\times \dots \times \mathcal{U}^{N}$ is said to be a Nash equilibrium for an $N$-player non-cooperative game if, for every $i \in {1,\dots,N}$ and $u \in \mathcal{U}^i$,
\begin{gather}
J_i(u^1,\dots,u^i,\dots,u^N) \le J_i(u^1,\dots,u^{i-1},u,u^{i+1},\dots,u^N),
\end{gather}
or equivalently
\begin{gather}
u^i=\mathop{\arg\min}_{u \in \mathcal{U}^i} J_i (u,u^{-i}).
\end{gather}
\end{definition}
Solving for an equilibrium can be challenging in dynamic games even with a small number of banks as each bank's strategy depends on all individual strategies of other banks. With a large number of decision makers involved, such problems become mathematically intractable. To tackle this issue, we use the mean-field game (MFG) methodology to obtain approximate ($\epsilon$-Nash) equilibrium solutions of the finite-player game. 

In $\epsilon$-Nash equilibria, an agent may have small incentives to unilaterally change its strategy. Hence, the strict requirement of Nash equilibrium, where no agent has any incentive to deviate from its strategy is relaxed. However, the incentive to deviate will not exceed $\epsilon$, where $\epsilon$ is typically a small value. This concept is expressed mathematically below. 

\begin{definition}[$\epsilon$-Nash property] An $N$-tuple of strategies $(u^1,\dots,u^{N}) \in \mathcal{U}^{1}\times \dots \times \mathcal{U}^{N}$ is said to be an $\epsilon$-Nash equilibrium solution for an $N$-player non-cooperative game if there exists an $\epsilon > 0$ such that for $i \in \{1,\dots,N\}$ and ${u} \in \mc{U}^i$,
\begin{gather}
    J_i(u^i,u^{-i}) \le J_i({u},u^{-i})+\epsilon,
\end{gather}
where $u$ is an admissible alternative strategy for agent-$i$.
\end{definition}
In the subsequent sections, we use the MFG methodology to obtain the set of borrowing and lending strategies that yields an $\epsilon$-Nash equilibrium. 

\section{Mean-Field Game Formulation}\label{section:MFG formulation}
MFG theory addresses a class of dynamic games involving a large number of agents who choose strategies to minimize their individual cost functionals. In such games, agents act in a non-cooperative fashion and aim to find their best strategy in response to the aggregate effect of the population modeled by the empirical distribution of states across the population of agents. This aggregate effect appears in the optimization problem through the dynamics and/or the cost functionals of agents. The general idea of the MFG methodology is that some useful simplifications occur in the limiting case with an infinite number of agents. MFG theory establishes the existence ofequilibria and characterizes them in terms of asymptotic strategies when the number of agents, $N$, in the system tends to infinity. More specifically, this set of asymptotic solutions yields a Nash equilibrium for the limiting game and an $\epsilon$-Nash equilibrium for the finite-player game, where the unilateral deviation incentives for each agent do not exceed a small value, $\epsilon$, in the latter case \cite{Lasry2007Meanfieldgames}. This method is referred to as the fixed-point (or top-down) approach. Another research direction in the MFG literature investigates the existence and characterization of a Nash equilibrium for the finite-player game, and studies its convergence to the Nash equilibrium of the limiting game \cite{Huang2006Largepopulationstochastic,Huang2007LargePopulationCostCoupledLQG}. This method is referred to as the direct (or bottom-up) approach which requires access to centralized
information on the states of all agents involved in the game and leads to a large-dimensional optimization problem. In some works in the context of interbank markets both the direct and fixed-point methods are used (see e.g. such as \cite{sun2022mean, Sun2018SystemicRiskInterbank, CarmonaSysRisk2015}). In this work, we employ the fixed-point approach, which leads to a lower-dimensional optimization problem and utilizes decentralized information, thereby reducing complexity in computation and implementation \cite{Minyi-Huang-Direct-Fixed-Point-App-2019}.

To proceed with the MFG formulation of the interbank market model, we define the mean-field of log-monetary reserves $\bar{x}_t$ and the mean-field of transactions with the central bank $\bar{u}_t$  in the limiting case as
\begin{gather}
    \bar{x}_t=\mathbb{E}[x^._t|\mc{F}^0_t], \label{MeanFieldState}\\
    \bar{u}_t=\mathbb{E}[u^._t|\mc{F}^0_t], \label{MeanFieldControl}
\end{gather}
where $x^._t$ and $u^._t$ denote, respectively, the log-monetary reserve and the borrowing and lending activities of a representative minor bank at equilibrium. If the limit exists, the mean-field terms are equivalent to the mathematical limit of the empirical averages at equilibrium as the number of banks $N$ goes to infinity as given by 
\begin{gather}
    \bar{x}_t=\lim_{N \to \infty}x^{(N)}_t=\lim_{N \to \infty}\frac{1}{N}\sum^N_{i=1}x^i_t,\allowdisplaybreaks\\
     \bar{u}_t=\lim_{N \to \infty}u^{(N)}_t=\lim_{N \to \infty}\frac{1}{N}\sum^N_{i=1}u^i_t.\label{MeanFieldControl-ave}
\end{gather}
Accordingly, we express the interbank model in the limiting case with (i) the major bank's optimal control problem, (ii) a representative minor bank's optimal control problem, and (iii) the mean-field equation as follows. \\
\noindent \underline{\textbf{(i) Major bank}} 
\begin{equation}
    dx_t^0 = a_0\left( \bar{x}_t-x_t^0\right)d t+u_t^0d t+\sigma_0 d W_t^0,\label{major_dyn_inf}
\end{equation}
\begin{equation}
\begin{aligned}
    J_0^\infty(u) = \mathbb{E} \Big[ \int_0^T \Big \{\frac{1}{2} \big(u_t^0\big)^2-q_0 u_t^0\big( \bar{x}_t- x_t^0\big)+ \frac{\epsilon_0}{2}\big( \bar{x}_t-x_t^0\big)^2 \Big \}d t
    +\frac{c_0}{2}\big( \bar{x}_T- x_T^0\big)^2 \Big].
\end{aligned}
\end{equation}
The information set of the major bank  is denoted by $\mathcal{F}^0=(\mathcal{F}^0_t)_{t \in[0,T]}$. It is generated by the sample paths of the state of the major bank. The admissible set $\mathcal{U}^0$ of control action for the major bank consists of all $\mathcal{F}^0$-adapted $\mathbb{R}$-valued processes such that $\mathbb{E} \Big[ \int_0^T (u_t^0)^2 dt \Big] < \infty$.\\

\noindent \underline{\textbf{(ii) A representative minor bank}}
\begin{gather}
    dx_t^i=a\left( \big(F \bar{x}_t+G x_t^0 \big)-x_t^i\right)d t+u_t^id t+\sigma d W_t^i, \label{minorDyn_infPop}\\
   J_i^\infty(u) = \mathbb{E}\Big[ \int_0^T \Big \{\frac{1}{2} \big(u_t^i\big)^2-q u_t^i\big(F \bar{x}_t+G x_t^0-x_t^i\big)+\frac{\epsilon}{2}\big(F \bar{x}_t+G x_t^0-x_t^i\big)^2 \Big \}d t
+\frac{c}{2}\big(F \bar{x}_T+G x_T^0-x_T^i\big)^2 \Big].\label{minor_dyn_inf}
\end{gather}
The information set of a representative minor bank $\mc{A}^i$ is denoted by $\mathcal{F}^i=(\mathcal{F}^i_t)_{t \in[0,T]}$. It is generated by the states of the major bank and the minor bank $\mc{A}^i$. The admissible set $\mathcal{U}^i$ of control action for the minor agent consists of all $\mathcal{F}^i$-adapted $\mathbb{R}$-valued processes such that $\mathbb{E} \Big[ \int_0^T (u_t^i)^2 dt \Big] < \infty$. This assumption guarantees that the optimal control problem is well-defined.\\

\noindent 
\underline{\textbf{(iii) Mean-field equation}}
We derive the mean-field dynamics as a function   of $\bar{u}_t$ by taking the expectation of the solution to \eqref{minorDyn_infPop} conditional on the information set $\mc{F}^0_t$. As a result the diffusion part disappears due to the independence of Brownian motions $\{w^0_t, w^i_t\}$ and the mean-field $\bar{x}_t$ satisfies
\begin{equation}\label{MeanFiledUbar}
    d \bar{x}_t=\left(a(F-1) \bar{x}_t+a G x_t^0+\bar{u}_t\right)dt,
\end{equation}
which characterizes the dynamics of the mass of minor banks as their number grows to infinity.

\section{Best-Response Transactions and Interbank Equilibria}\label{section:BestRespnseTranscation}
The MFG methodology yields a set of results, which are summarized by the following theorems. Detailed derivations can be found in \ref{method}. First, we present the set of optimal transaction strategies. Next, we demonstrate that this set leads to a Nash equilibrium for the limiting interbank model, and an $\epsilon$-Nash equilibrium for the finite-player market.

\begin{theorem}[Best-Response Transactions] \label{thm:Best_resp}For the limiting interbank market model given by \eqref{major_dyn_inf} - \eqref{MeanFiledUbar} and subject to the clearing conditions \eqref{ClearingCondition1} - \eqref{ClearingCondition2}, the optimal borrowing and lending strategies for the major bank and a representative minor bank, and the mean-field equation are given by\\
\noindent \underline{\textbf{(i) Major bank}}: 
\begin{itemize}
\item Optimal strategy
\begin{equation}\label{MajorOptimaControl_thm}
    u_t^{0,*}=\big(q_0-\phi^0_t\big) \big(  \bar{x}_t - x_t^0\big),
\end{equation}
with the coefficient $\phi^0_t$ satisfying
\begin{equation}\label{MajorRaccati_thm}
\begin{split}
    \dot{\phi}_t^0&=2\left((a_0+q_0) +G\big(a+q-\phi_t \big)\right)\phi_t^0-\big( \phi_t^0\big)^2 +\epsilon_0 -q_0^2, \quad
    \phi_T^0=- c_0.
\end{split}
\end{equation}
\end{itemize}
\noindent \underline{\textbf{(ii) Representative minor bank}}:
\begin{itemize}
    \item Optimal strategy
    \begin{equation}\label{MinorOptimaControl_thm}
    u_t^{i,*}=\big(q-\phi_t \big) \big[\big(F \bar{x}_t +G x_t^0\big)-x_t^i\big],
\end{equation}
    with the coefficient $\phi_t$ satisfying 
    \begin{equation}\label{MinorRaccati_thm}
\begin{split}
    \dot{\phi}_t=2(a+q) \phi_t -\big( \phi_t\big)^2+\epsilon -q^2, \quad \phi_T=- c.
\end{split}
\end{equation}
\end{itemize}
\noindent \underline{\textbf{(iii) Mean-field equation}}:
\begin{gather}\label{MeanFieldEquation} 
    d \bar{x}_t = \big(a+q-\phi_t\big) \Big( (F-1) \bar{x}_t +G x_t^0\Big)dt.
\end{gather}
\end{theorem}
\proof See \ref{method}. \rulex

We note that we are interested in an equilibrium for the original finite-population interbank market model described by \eqref{MajorDynamicsOriginal}-\eqref{MinorCostOriginal}. Now we connect the obtained solutions for the limiting model to the finite-population model through the notion of $\epsilon$-Nash equilibrium.
\begin{theorem}[Interbank Equilibria] Consider the best-response transaction strategies for the major bank and a representative minor bank characterized respectively by \eqref{MajorOptimaControl_thm}-\eqref{MajorRaccati_thm} and  \eqref{MinorOptimaControl_thm}-\eqref{MinorRaccati_thm}.
\begin{itemize}
    \item[(i)] For the limiting interbank market described by \eqref{major_dyn_inf}-\eqref{MeanFiledUbar}, the set of best-response transaction strategies $U^\infty=\{u^0, u^1,\dots,u^{\infty}\}$ yields a Nash equilibrium.
    \item[(ii)] For the finite-population interbank market described \eqref{MajorDynamicsOriginal}-\eqref{MinorCostOriginal}, the set of best-response strategies $U^{[N]}=\{u^0,u^1,\dots,u^{N}\}$ yields an $\epsilon$-Nash equilibrium.
\end{itemize}
\end{theorem}

\proof Given that all banks are following the strategies from $U^\infty$, the mean-field satisfies \eqref{MeanFieldEquation}. Now if a minor bank unilaterally deviates from $U^\infty$, as individually it has a negligible impact, this deviation does not affect the mean-field value and its characterization. Hence, the minor bank seeks an optimal strategy in response to the same mean-field as before. This yields to the strategy specified by \eqref{MinorOptimaControl_thm}-\eqref{MinorRaccati_thm}. Hence the minor bank cannot benefit by deviating unilaterally. A similar reasoning can be used for the unilateral deviation of the major bank from $U^\infty$. In this case still the mean-field satisfies \eqref{MeanFieldEquation}, where the value of $x^0_t$ is updated. This results in the same optimal control law for the major bank. Hence $U^\infty$ forms a Nash equilibrium for the limiting interbank market model \eqref{major_dyn_inf}-\eqref{MeanFiledUbar} (for further details see e.g. \citep{Huang2010LargePopulationLQGGames}.).

The interbank model considered is a special case of LQG mean-field games with one major agent and a large population of minor agents. Hence, the proof of $\epsilon$-Nash property follows from the existing results in the literature, see e.g. 
\citep{Huang2010LargePopulationLQGGames,Carmona2016probabilisticapproachmean}.   \rulex

Next, we investigate the risk of individual bank's default and a systemic event for the interbank market in equilibrium. 

\section{Individual Default and Systemic Risk} \label{section:Individual Default and Systemic Risk in Interbank Transactions}
Banks act as financial intermediaries between borrowers and lenders. However, they face a risk of default if they have insufficient available reserves to make the required payments to the lenders. 
The risk of default faced by a bank is influenced not only by its own reserves but also by the conditions of other banks in the market. This interdependence among banks arises from the interbank transaction market and poses a threat to the entire financial system, known as systemic risk. In other words, systemic risk refers to the potential risk of a financial system's collapse as a result of such interconnections.

More specifically, we follow the definition of a default proposed in \citet{CarmonaSysRisk2015} which corresponds to  the scenario where the log-monetary reserve of a bank goes below a specific value called the default threshold. It could also be related to ``reserve requirements'' established by the financial regulator. The systemic event, or system default, occurs when the market state, as measured by the market-level log-monetary reserves, falls in the default region. We redefine the market state when a major bank is present as a linear combination of the major bank's log-monetary reserve and the average log-monetary reserve of the minor banks. The weights used in this combination indicate the relative size of the two forces operating in the market - the major bank and the mass of minor banks. This combination is represented as $F{x}_t^{(N)} + Gx^0_t$ in the finite-population interbank model and as $F\bar{x}_t + Gx^0_t$ in the limiting model. In this section, we provide the definition of default probabilities for the finite-population model. The corresponding definitions for the limiting model can be obtained by employing the limiting log-monetary reserves and market state.

Let us denote the default threshold by $D$. 
The default probability of bank-$i,\, i \in \{0, 1, \dots\},$ is defined as
\begin{equation}
p_i = \mb{P} \Big(\text{default of bank-$i$}\Big) = \mb{P} \Big(\min_{t\in[0,T]}\ (x_t^i)\ \le \ D \Big).\label{def_D_risk}
\end{equation}
The probability of systemic event, or  systemic risk, is defined by 
\begin{equation} 
p_{SE}=\mb{P}(\text{systemic event})=\mb{P} \ \Big(\min_{t\in[0,T]}\ \big( F {x}_t^{(N)} +G x^0_t \big) \ \le \ D \Big).
\end{equation}

In this work, we are interested in the role played by the major bank on the default risk $p_i$ of a representative small bank-$i,\, i \in \{1,2,\dots\}$ and on the systemic risk $p_{SE}$. As a benchmark case, we use the results in absence of a major bank as obtained by  \citet{CarmonaSysRisk2015}, \citet{Fouque2013StabilityModelInterbank}, and \citet{Fouque2013SystemicRiskIllustrated}. To understand the role played by the major bank, we also study the  probabilities of each event conditional on the major bank having defaulted or not. Formally,  for $i \in \{1,2, \dots\}$, we are interested in
$\textit{p}_{i|MD} =\mb{P} \ \Big(\min_{t\in[0,T]}\ x_t^i\ \le \ D \big| \min_{t\in[0,T]}\ x_t^0\ \le \ D \Big)$,
    $\textit{p}_{i|MS} =\mb{P} \ \Big(\min_{t\in[0,T]}\ x_t^i\ \le \ D \big| \min_{t\in[0,T]}\ x_t^0\ > \ D\Big)$,
and the conditional systemic risks $\textit{p}_{SE|MD} =\mb{P} \ \Big(\min_{t\in[0,T]}\! \big(F{x}_t^{(N)} +G x^0_t \big) \le \ D \big|\min_{t\in[0,T]} x_t^0\ \le \ D\Big)$, 
 $\textit{p}_{SE|MS} =\mb{P} \ \Big(\min_{t\in[0,T]}\!\big(F{x}_t^{(N)} +G x^0_t \big) \le \ D\, \big|\min_{t\in[0,T]}x_t^0\ > \ D\Big)$.
 Note that the law of total probability implies $p_i = \left({p}_{i|MD}-{p}_{i|MS}\right)p_0 + {p}_{i|MS}$ and $p_{SE} = \left({p}_{SE|MD}-{p}_{SE|MS}\right)p_0 + {p}_{SE|MS}$
 where, from \eqref{def_D_risk}, $p_0$ denotes the major bank's default risk. 
 
 Due to the mathematical challenges involved in the analytical calculation of default probabilities, we employ Monte Carlo simulations to determine these probabilities in the next section. 

\section{Numerical experiments}\label{section:Numerical Experiments} 
In this section, we perform numerical experiments to obtain insights into the role played by major banks and market frictions on the financial system's stability. More specifically, we explore the influence of the relative size $G$ and the mean reversion rate $a$.  
These simulations are performed separately for both finite and infinite populations of minor banks.
For the finite-population case, we consider a setup where there are $10$ minor banks and one major bank  and perform 50,000 simulations for various settings. The realizations of the stochastic processes that model uncertainty are the same across considered scenarios (e.g. for different $G$) throughout the simulations.
Remark that the strategies employed by banks in the finite population correspond to the limiting strategies, where the mean-field $\bar{x}_t$ and the limiting log-monetary reserves are, respectively, replaced by the empirical average $x^{(N)}_t$ of the log-monetary reserves of minor banks and the log-monetary reserves specific to  the finite-population. 
Hence, the equilibrium of a large population game is here approximated by the Nash equilibrium of the limiting game when the number of agents goes to infinity\footnote{We investigate the quality of this approximation in Online Appendix \ref{app:approxquality}.}. \citep{CarmonaSysRisk2015} shows that the financial implications using approximate Nash equilibrium are identical to the ones where the exact Nash equilibrium is used.

We choose the default threshold $D=-0.65$,\footnote{This value, commonly employed in the literature for similar interbank models, corresponds to the $1\%$ quantile of the distribution of log-monetary reserve of small banks for the case where $a=F=G=1$.} and assume all banks remain in the system and continue to lend and borrow until the end of time even if they have crossed the default threshold. We perform Mont Carlo simulations and report estimates of the probabilities of interest, outlined in Section \ref{section:Individual Default and Systemic Risk in Interbank Transactions} and denoted by $\bar{p}_i, \bar{p}_{i|MD}, \bar{p}_{i|MS}, \bar{p}_{SE}, \bar{p}_{SE|MD}, \bar{p}_{SE|MS}$, for the cases where the relative size $G$ and the mean reversion rate $a_0$ of the major bank change. Specifically, the notation $\bar{p}$ refers to the estimate of the corresponding probability $p$ obtained from the Monte-Carlo simulations.  We only report the main results here and many additional simulation results are presented in the Online Appendix.

We note that due to limited interbank trading data for individual banks, model tuning becomes impractical. Therefore, we rely on interpreting the results based on the relative changes in these probabilities rather than their absolute values. This approach allows us to gain insights into the behavior of the interbank system without requiring precise calibration of the model.

\subsection{Size of the Major Bank}\label{section:Numerical_finite_G} 
Each minor bank tracks the average log-monetary reserve in the market (or the market state) given by $(F {x}_t^{(N)} +G x_t^0)$ in the finite population and by $(F \bar{x}_t +G x_t^0)$ in the infinite population, respectively. This market state is a weighted average of the major bank's log-monetary reserve and the average level of log-monetary reserves across small banks. The parameters $G$ and $F$ denote, respectively, the relative sizes of major bank and the mass of small banks. We perform simulations of the model for different values of $G \in \{0.1, 0.2, 0.3,\dots, 0.9\}$.
As the relative size of the major bank increases, all trajectories evolve more closely together. Furthermore, the larger the major bank the faster it brings the system towards a systemic default.

The total and conditional probabilities of a systemic event for the finite-population case  are shown in \cref{RegressionEstimate_FG_Finite:SystemicRisk}.  Our findings reveal that systemic risk generally increases in the presence of a major bank, except when the major bank is of small relative size ($G=0.1$). Additionally, the level of systemic risk is further exacerbated by the size of the major bank. Notably, the conditional systemic risk ($\bar{p}_{SE|MS}$) given that the major bank does not default is close to zero, indicating that a stable major bank significantly enhances system stability and helps mitigate systemic risk. However, the default of the major bank significantly amplifies systemic risk by attracting numerous small banks into the default zone. Therefore, the presence of a stable major bank in the interbank market has predominantly positive effects on systemic risk. However, it is crucial to note that the market stability provided by the major bank might not be sufficient to generate a positive net effect on systemic risk unless the large bank is relatively small compared to the overall interbank market or is subject to adequate regulation to mitigate the risk of default. These findings hold true in the infinite population setting as well.

\begin{table*}[h]
\centering
\scalebox{0.9}{
\begin{tabular}{ccccc}
\hline
 & No Major Bank  & \multicolumn{3}{c}{With a Major Bank}                                                                     \\ \cline{3-5} 
                
                   &                                & \multicolumn{1}{c}{}                       & \multicolumn{1}{c}{Non-defaulting Major} & Defaulting Major \\
                 G  &                     \multicolumn{1}{c}{$\bar{p}_{SE}$}           & \multicolumn{1}{c}{$\bar{p}_{SE}$}        & \multicolumn{1}{c}{$\bar{p}_{SE|MS}$}     & $\bar{p}_{SE|MD}$ \\ \hline
0                                        & 0.0348                         & \multicolumn{1}{c}{-}                              & \multicolumn{1}{c}{-}                    & -                                     \\ 
0.1                                      & -                              & \multicolumn{1}{c}{0.0321}                         & \multicolumn{1}{c}{0.0021}               & 0.0711                                \\ 
0.2                                      & -                              & \multicolumn{1}{c}{0.0613}                         & \multicolumn{1}{c}{0.0003}               & 0.1433                                \\ 
0.5                                      & -                              & \multicolumn{1}{c}{0.1838}                         & \multicolumn{1}{c}{0.0001}                    & 0.4588                                \\ 
0.7                                      & -                              & \multicolumn{1}{c}{0.2622}                         & \multicolumn{1}{c}{0}                    & 0.6774                                \\ 
0.9                                      & -                              & \multicolumn{1}{c}{0.3379}                         & \multicolumn{1}{c}{0}                    & 0.8934                                \\ \hline
\end{tabular}}
\caption{Estimated probability of systemic event in the finite-population model for the cases (from left to right) with (i) no major bank, (ii) total default probability ($\bar{p}_{SE}$) with a major bank, (iii) conditional default probability ($\bar{p}_{SE|MS}$) with a non-defaulting major bank, and (iv) conditional default probability ($\bar{p}_{SE|MD}$) with a defaulting major bank.}
\label{RegressionEstimate_FG_Finite:SystemicRisk}
\end{table*}

\begin{figure}
\subfigure[]{
\begin{minipage}[t]{\textwidth} 
\centering
    \includegraphics[width=0.48\textwidth]{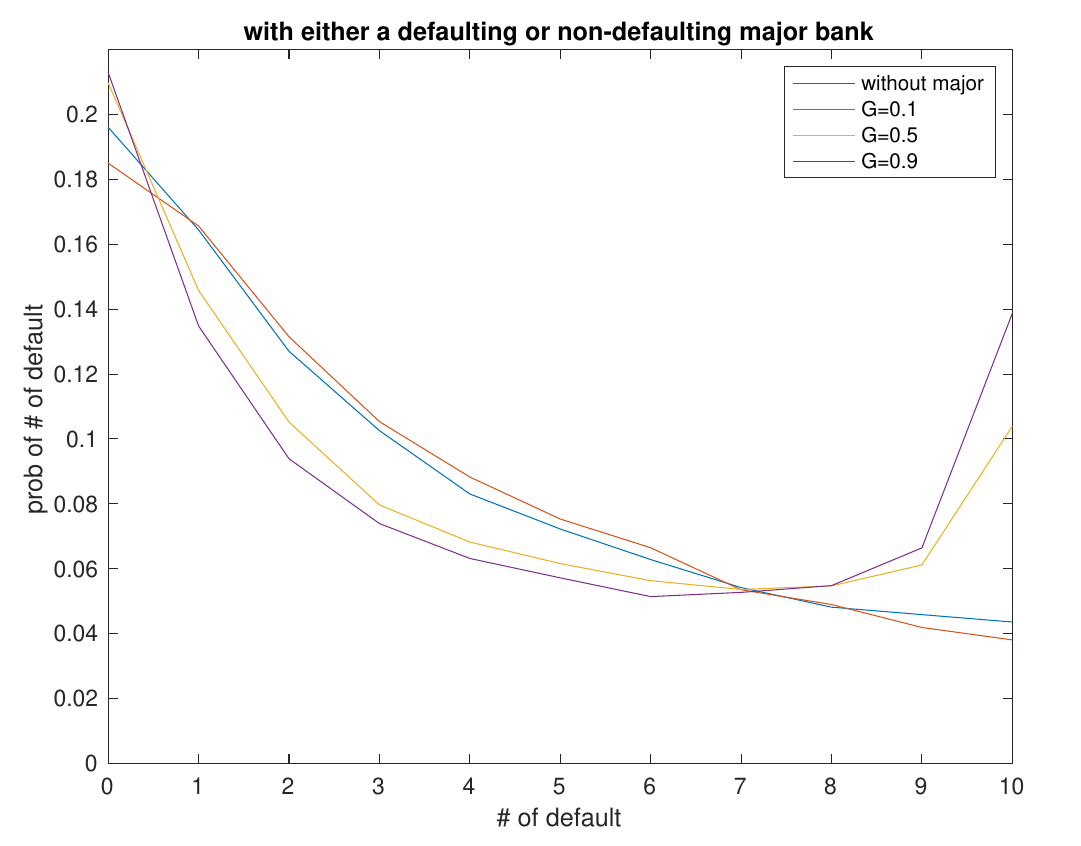}
    \end{minipage}}
\vspace{10pt}
\subfigure[]{
\begin{minipage}[t]{.48\textwidth} 
\includegraphics[width=\textwidth]{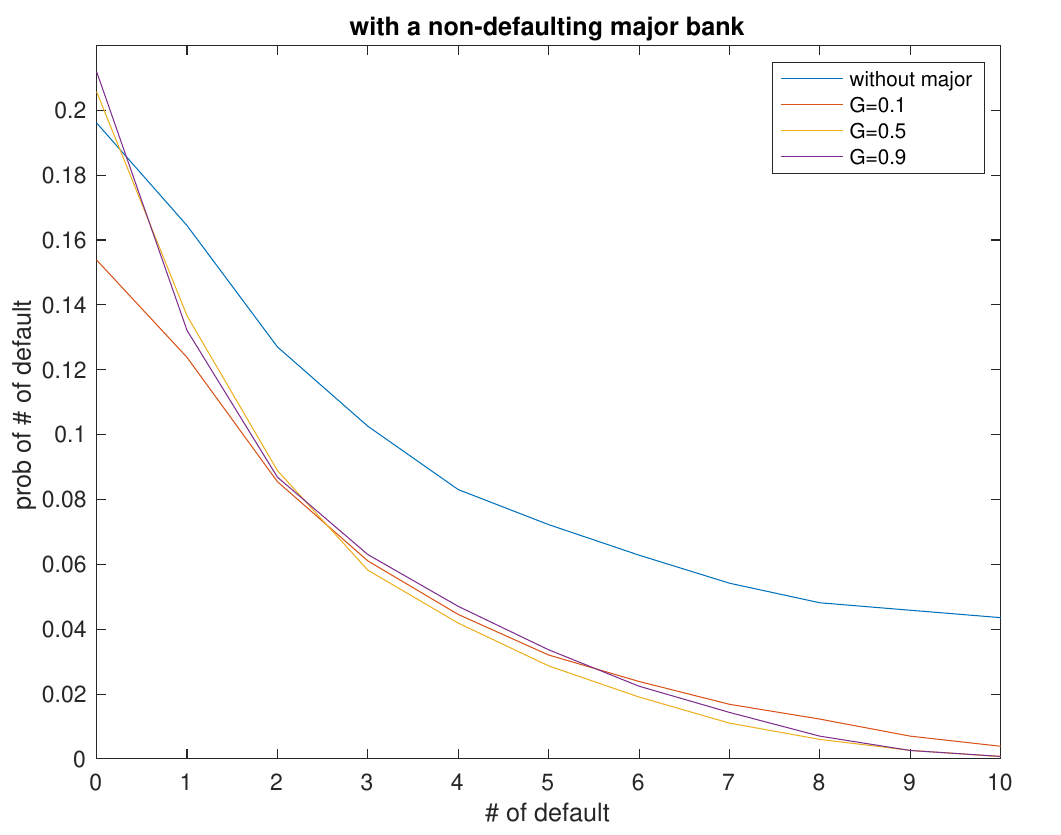}
\end{minipage}}
\hfill 
\subfigure[]{
\begin{minipage}[t]{.48\textwidth} 
\includegraphics[width=\textwidth]{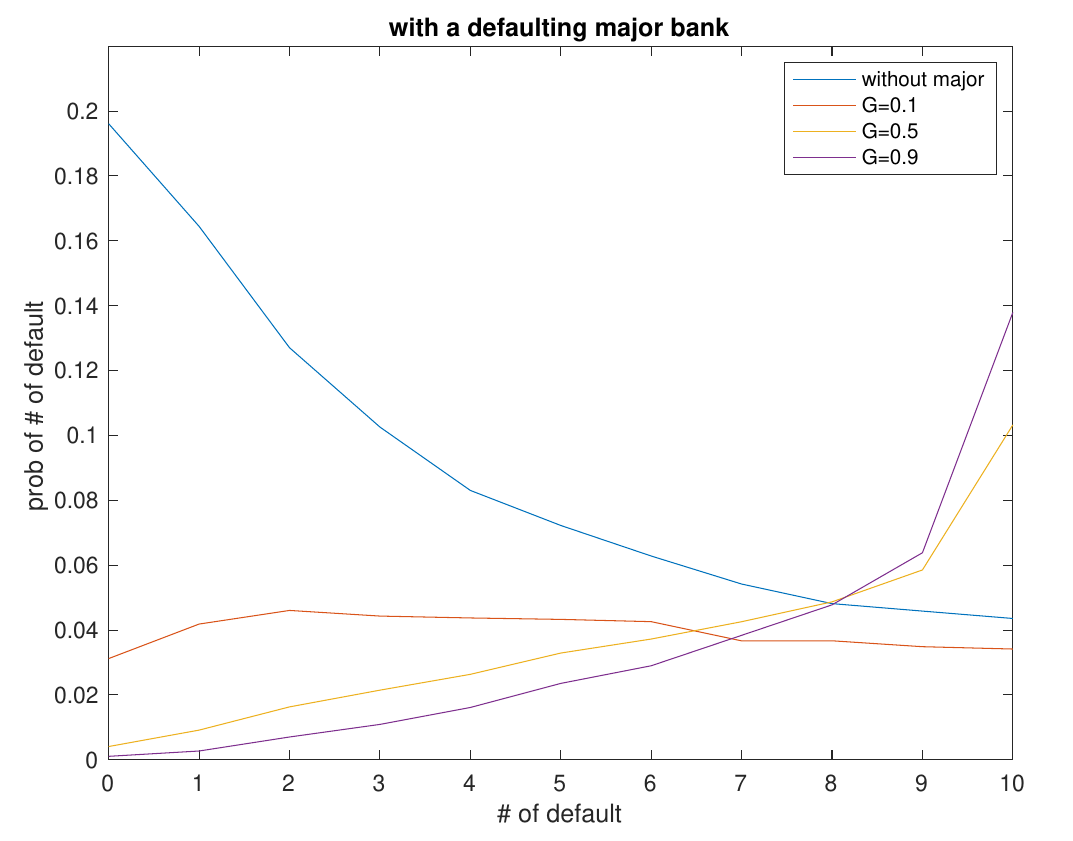}
\end{minipage}}
\caption{Loss distribution ($a=5$): (a) total loss distribution for minor banks, (b) loss distribution for minor banks conditional on the major bank not default, and (c) loss distribution for minor banks conditional on the major bank default.}\label{fig:LossDistribution_FG}
\end{figure}

We plot the loss distribution of minor banks in the finite population setting in \cref{fig:LossDistribution_FG}. It corresponds to the probability distribution of the number of defaults. In panel (a), the tail of the distribution gets fatter as we increase the relative size of the major bank. This result means that the probability of extreme events, i.e. either a large number of minors go to default together or no minor bank defaults, increases with $G$. Panel (b)  shows that having a stable major bank improves the stability of the system. As the relative size of the major agent increases the loss distribution remains almost the same except in the left tail, i.e. the probability of the extreme event where no bank ends up in default increases with $G$. Panel (c) shows that if the major bank defaults, the right tail of the loss distribution becomes much fatter as  $G$ increases, meaning that the probability of the extreme event where all banks wind up in default increases. 

\subsection{Role of Market Frictions}\label{section:Numerical_finite_a}
We now assume that the major bank and the mass of minor banks are of the same size ($F=G=0.5$) and examine the impact of reducing market frictions, by increasing the mean-reversion rate $a$. Recall that we interpret $a_0$ and $a$ as inverse measures of market frictions. A smaller value of $a$ means larger market friction for all banks, since $a_0$ and $a$ parameters are jointly determined with $F$ and $G$, as already established when discussing the market clearing condition \eqref{ClearingCondition1}-\eqref{ClearingCondition2}. We therefore interpret larger values for $a_0$ and $a$ as indicative of the degree to which banks rely on the interbank market to satisfy their financial obligations.  

In this case, we have $a_0=0.5a$. A higher mean-reversion rate translates into a higher frequency of lending and borrowing activities. Thus, the major bank trades at a lower frequency than a representative minor bank given the same distance from their respective tracking signal, respectively, $\bar{x}_t$ and $0.5(x^0_t+\bar{x}_t)$. This could be due to some market frictions and conditions  as explained in \cref{section:Market Clearing Condition}. To investigate the impact of the mean reversion rates on the system we consider $a \in \{1,2,\dots,10\}$.

The results for systemic risk are summarized in \cref{RegressionEstimate_a_Finite:SystemicRisk}.
In the absence of a major bank, the estimated probability of a systemic event is found to be small. Moreover, the level of systemic risk appears to be relatively unaffected by the mean-reversion rate $a$ in this context. On the contrary, in the presence of a major bank, the systemic risk $\bar{p}_{SE}$ shows an increasing trend with respect to $a$. To gain further insights, we can examine the conditional systemic risks. When the major bank remains stable, assuming $F=G=0.5$, the systemic risk $\bar{p}_{SE|MS}$ decreases to zero. However, in the event of the major bank failure, the systemic risk $\bar{p}_{SE|MD}$ significantly worsens, and its magnitude further amplifies with increasing $a$. These findings hold true for the infinite population case as well. 

\begin{table*}[h]
\centering
\scalebox{0.9}{
\begin{tabular}{ccccc}
\hline
 & No Major Bank  & \multicolumn{3}{c}{With a Major Bank}                                                                     \\ \cline{3-5} 
                
                   &                                & \multicolumn{1}{c}{}                       & \multicolumn{1}{c}{Non-defaulting Major} & Defaulting Major \\
                 $a$  &                     \multicolumn{1}{c}{$\bar{p}_{SE}$}           & \multicolumn{1}{c}{$\bar{p}_{SE}$}        & \multicolumn{1}{c}{$\bar{p}_{SE|MS}$}     & $\bar{p}_{SE|MD}$ \\ \hline
1                                        & 0.0371                         & \multicolumn{1}{c}{0.1755}                         & \multicolumn{1}{c}{0}                    & 0.3957                                \\ 
3                                        & 0.0367                         & \multicolumn{1}{c}{0.1795}                         & \multicolumn{1}{c}{0}                    & 0.4238                                \\ 
5                                        & 0.0361                         & \multicolumn{1}{c}{0.1848}                         & \multicolumn{1}{c}{0}                    & 0.4585                                \\ 
7                                        & 0.0362                         & \multicolumn{1}{c}{0.1870}                         & \multicolumn{1}{c}{0}                    & 0.4856                                \\ 
10                                       & 0.0376                         & \multicolumn{1}{c}{0.1932}                         & \multicolumn{1}{c}{0}                    & 0.5278                               \\ \hline
\end{tabular}}
\caption{Estimated probability of systemic event in finite-population model for the cases (from left to right) with (i) no major bank, (ii) total default probability ($\bar{p}_{SE}$) with a major bank, (iii) conditional default probability ($\bar{p}_{SE|MS}$) with a non-defaulting major bank, and (iv) conditional default probability ($\bar{p}_{SE|MD}$) with a defaulting major bank.}
\label{RegressionEstimate_a_Finite:SystemicRisk}
\end{table*}

\section{Concluding Remarks}\label{section:concluding remarks}
In this paper, we have conducted an analysis of the impacts of a large bank on smaller banks and systemic risk through the use of a MFG model. Our numerical results indicate that the presence of a major bank can have both stabilizing and destabilizing effects on the financial system. On the one hand, the major bank can provide additional stability to the system by allowing the smaller banks to better coordinate and reduce uncertainty. On the other hand, it also introduces increased connectivity, which can lead to larger consequences in the event of a default. The net result of these opposing effects depends on the market size of the large bank. As the market size of the large bank increases, the negative externality of added systemic risk tends to offset the stability gains provided by the coordination channel. However, it is worth noting that the large bank always seems to provide a net benefit to the system as long as it does not default.

Additionally, our results suggest that not only the size of the large bank is important, but also the level of market friction that limits trading frequencies. The speed of trading has two opposing effects on systemic risk that are highly dependent on the state of the major bank. While a higher trading frequency can reduce systemic risk in the presence of a stable major bank, it can also dramatically increase systemic risk if the major bank experiences financial distress, as all of the smaller banks use it as a common market signal.

These findings have important implications for understanding the dynamics of interbank transaction markets and the role of large banks within them. They suggest that policies and regulations aimed at improving the stability of these markets should focus on ensuring the stability of large banks, possibly through the use of higher capital requirements, limiting the size of these banks, and monitoring market conditions and regulating trading frequencies accordingly. However, it is important to recognize that the model used in this study has certain limitations, including its relative simplicity, which may not fully capture the complexity of interbank activities. Further research is needed to more fully understand the impact of various parameters on the system and to study systemic risk and the behavior of a major bank in a ``too big to fail" policy context. It may also be useful to extend the model to consider multiple groups of minor banks with different characteristics and risk sensitivity, or several major banks of different sizes.

\bibliographystyle{elsarticle-num-names} 
\bibliography{refrences.bib}

\appendix
\renewcommand{\thesection}{\Alph{section}}
\section{Methodology}\label{method}
In this section we present each step of the solution methodology in detail to derive the optimal trading strategy for the major bank and a representative minor bank $\mc{A}^i$ for the limiting interbank model given by \eqref{major_dyn_inf} - \eqref{MeanFiledUbar}. We note that we cannot directly use the results derived by \citet{Huang2010LargePopulationLQGGames} and \citet{Firoozi2020ConvexanalysisLQG} for MFG models with a major agent. This is because, for the purpose of our work, we are interested in deriving the optimal transaction rates in terms of the difference between the market state (or the market's average log-monetary reserve) and the bank's log-monetary reserves (i.e. $(x_t^{(N)} - x_t^0)$ for the major bank or $(F x_t^{(N)} +G x_t^0 -x_t^i)$ for a representative small bank.). We begin by addressing the optimization problem of the major agent. Motivated by the convex analysis approach introduced by \citet{Firoozi2020ConvexanalysisLQG}, we introduce a small perturbation to the major bank's strategy and study its propagation throughout the entire economy. By subsequently calculating the G\^ateaux derivative of the major bank's cost functional and setting it to zero, we derive the necessary and sufficient optimality condition for the major bank's transaction strategy.
To further elucidate the major bank's optimal strategy, we need to characterize the mean-field behavior of the log-monetary reserves. Hence, we delve into the analysis of an individual representative minor bank's problem. Employing a similar variational approach, we solve the optimization problem specific to the minor bank and accordingly derive the mean field dynamics. Then, we return to the major bank's problem and conclude the analysis by deriving an explicit representation of its optimal strategy.

\subsection{Major Bank's Problem}

\vspace{0.2cm}
\noindent \underline{\textbf{Step (i)}}: We perturb the strategy of the major bank by $\delta_0$ in the direction $\omega^0\in\mc{U}^0$.
The dynamics of the major bank's log-monetary reserve subject to the perturbed strategy $u^0_0+\delta_0\omega^0$ is given by
\begin{equation}
    d x_t^{0,\delta_0}=a_0\big(\bar{x}_t^{\delta_0}-x_t^{0,\delta_0}\big)d t+\big(u_t^0+\delta_0\omega^0\big)d t+\sigma_0 d W_t^0.
\end{equation}

\noindent \underline{\textbf{Step (ii)}}: We investigate the impact of the major bank's perturbed strategy on its own and each minor bank's log-monetary reserve in order to obtain the resulting perturbed mean-field $\bar{x}_t^{\delta_0}$.

The perturbed log-monetary reserve of a representative minor banks $x^{i,\delta_0}$ is given by
\begin{equation}\label{MinorDynamicsPerturbed_0}
    d x_t^{i,\delta_0}=a\left(\big(F \bar{x}_t^{\delta_0}+G x_t^{0,\delta_0}\big)-x_t^{i,\delta_0}\right)d t+u_t^id t+\sigma d W_t^i.
\end{equation}
Subsequently the perturbed mean-field  $\bar{x}_t^{\delta_0}$ is obtained by taking the conditional expectation of $x^{i,\delta_0}$ given $\mc{F}^0_t$ which satisfies
\begin{equation}\label{MeanFiledPerturbed_0}
    d \bar{x}_t^{\delta_0}=\left(a(F-1) \bar{x}_t^{\delta_0}+a G x_t^{0,\delta_0}+\bar{u}_t\right)dt.
\end{equation}

By examining equations \eqref{MinorDynamicsPerturbed_0} and \eqref{MeanFiledPerturbed_0}, we gain valuable insights into the impact of perturbing the major bank's strategy on both the minor banks and the overall system. This understanding highlights the benefits offered by the convex analysis approach, as discussed in \citep{Firoozi2020ConvexanalysisLQG}. The perturbation in the transaction strategy of the major bank with the central bank has a direct impact on its own log-monetary reserves. Furthermore, owing to its unique position in the market, the perturbation in the log-monetary reserve of the major bank directly influences the log-monetary reserves of the minor banks and indirectly affects the mean-field of log-monetary reserves. Subsequently, the perturbed mean-field influences the evolution of the major bank's log-monetary reserves. Thus, we can observe the interconnectedness and mutual influence between the major bank, minor banks, and the overall system as a result of perturbations in the major bank's transaction strategy.

\vspace{0.1cm}
\noindent
\underline{\textbf{Step (iii)}}: In line with the methodology established in \citep{Huang2010LargePopulationLQGGames} for MFG models featuring a major agent, we construct the major bank's extended dynamics by combining its individual dynamics with that of the mean field. This approach allows us to capture the interaction and interdependence between the major bank and the collective behavior represented by the mean field. The major bank's extended state $(X_t^{0,\delta_0})^\intercal = [
(x_t^{0,\delta_0})^\intercal\,\,\,
(\bar{x}_t^{0,\delta_0})^\intercal]$ satisfies
\begin{equation}\label{MajorExtendDynmaic}
d X_t^{0,\delta_0}
=\left(\Tilde{A_0}X_t^{0,\delta_0}+\mb{B}_0u^0_t+\Tilde{B_0}\bar{u}_t+\delta_0\mb{B}_0\omega^0 \right)d t + \Sigma_0d \overline{W}_t^0,
\end{equation}
where
\begin{equation*}\label{MajorExtendDynmaicMatrices}
 \tilde{A}_0 = \left[ \begin{array}{cc}
-a_0 & a_0 \\
a G &  a(F-1)
\end{array} \right]\!,
\quad \mb{B}_0=\left[ \begin{array}{c} 1 \\ 0  \end{array}\right]
\!,
\quad
 \tilde{B}_0 = \left[ \begin{array}{c} 0 \\ 1 \end{array}\right]\!, 
\quad
\Sigma_0 = \left[ \begin{array}{cc}
\sigma_0 & 0 \\
0 &  0
\end{array} \right]\!,\quad \overline{W}_t^0 = \left[ \begin{array}{c}W^0_t\\0\end{array} \right].
\end{equation*}
Moreover, the major bank's cost functional in terms of the extended state $X_s^{0,\delta_0}$ is given by
\begin{gather}\label{MajorExtCostOriginal}
J^0(u^0+\delta_0\omega^0) = \tfrac{1}{2}\mb{E} \bigg [ \int_0^T \Big \{(X_s^{0,\delta_0})^\intercal \mb{Q}_0 X_s^{0,\delta_0} + 2(X_s^{0,\delta_0})^\intercal \mb{N}_0 (u_s^0 +\delta_0 \omega^0_s )+ (u^0_s + \delta_0 \omega^0_s)^2\Big \} dt + (X_T^{0,\delta_0})^\intercal \mb{G}_0 X_T^{0,\delta_0}  \bigg],\\
 \mb{Q}_0 = \left[ \begin{array}{cc}
\epsilon_0 & -\epsilon_0  \\
-\epsilon_0   &  \epsilon_0  
\end{array} \right]\!,
\quad \mb{N}_0=\left[ \begin{array}{c} q_0 \\ -q_0   \end{array}\right]
\!,
\quad
 \mb{G}_0 = \left[ \begin{array}{cc}
c_0 & -c_0 \\
-c_0   &  c_0 
\end{array} \right]\!.\label{MajorExtendCostMatrics}
\end{gather}

\noindent 
It is worth noting that the unperturbed extended dynamics and cost functional for the major bank can be obtained by setting the perturbation in \eqref{MajorExtendDynmaic}-\eqref{MajorExtCostOriginal} to zero. This leads to the following model:
    \begin{equation}\label{MajorExtendStateDynamicOriginal}
        d X_t^0=\left(\Tilde{A_0}X_t^0+\mb{B}_0u^0_t+\Tilde{B_0}\bar{u}_t\right)d t + \Sigma_0d \overline{W}_t^0,
    \end{equation}
    \begin{equation}\label{MajorExtendCostPerturbed}
     J^0(u^0) = \tfrac{1}{2}\mathbb{E} \bigg [ \int_0^T \Big \{ \big(X_s^{0}\big)^\intercal \mb{Q}_0 X_s^{0} + 2\big(X_s^{0}\big)^\intercal \mb{N}_0 \big(u_s^0  \big)+ \big(u^0_s\big)^2\Big \} dt + \big(X_T^{0}\big)^\intercal \mb{G}_0 X_T^{0}  \bigg].
     \end{equation}
Now we aim to characterize the mean-field of transaction strategies $\bar{u}_t$ appearing in the major bank's dynamics given by \eqref{MajorExtendDynmaic}. For this purpose, we look into the problem of a representative small (minor) bank.

\subsection{Minor Bank's Problem}

\vspace{0.2cm}
\noindent \underline{\textbf{Step (i)}}: We perturb a representative minor bank's strategy by $\delta_i$ in the direction $\omega^i\in\mc{U}^i$. The perturbed state $x_t^{i,\delta_i}$ satisfies
\begin{equation}\label{MinorDynamicsOriginal_i} 
    d x_t^{i,\delta_i}=a\left(\big(F\bar{x}_t^{\delta_i}+G x_t^{0,\delta_i}\big)-x_t^{i,\delta_i}\right)d t+\big(u_t^i+\delta_i\omega^i\big) d t+\sigma d W_t^i.
\end{equation}

\noindent \underline{\textbf{Step (ii)}}: We obtain the perturbed mean field by taking the conditional expectation of $x_t^{i,\delta_i}$ given $\mc{F}^0_t$. The perturbed mean-field $\bar{x}^{\delta_i}$ in this case satisfies
\begin{equation}\label{MeanFieldPerturbed_i}
d \bar{x}_t^{\delta_i}=\left(a(F-1) \bar{x}_t^{\delta_i}+a G x_t^{0,\delta_i}+\bar{u}_t\right)d t.
\end{equation}
We observe that the perturbation of a minor bank's strategy propagates throughout the  system in a different fashion than that of the major bank. From \eqref{MinorDynamicsOriginal_i} and \eqref{MeanFieldPerturbed_i}, a perturbation in a small bank's trading activity with the central bank affects its own log-monetary reserve. However, due to the negligible impact of one minor bank, the mean-field and major bank are not effected by this perturbation resulting in $\bar{x}_t^{\delta_i}=\bar{x}_t$ and $x_t^{0,\delta_i}=x_t^0$.

\vspace{0.1cm}
\noindent \underline{\textbf{Step (iii)}}: We extend the minor bank's state to include the major bank's log-monetary reserve and the mean field as in
\begin{equation}\label{MinorExtendStateMatrices}
X_t^{i,\delta_i} = \begin{bmatrix}
x_t^{i,\delta_i}\\
x_t^{0,\delta_i}\\
\bar{x}_t^{0,\delta_0}
\end{bmatrix}
= \begin{bmatrix}
x_t^{i,\delta_i}\\
x_t^{0}\\
\bar{x}_t
\end{bmatrix}
= \begin{bmatrix}
x_t^{i,\delta_i}\\
X_t^{0,\delta_0}
\end{bmatrix}.
\end{equation}
Subsequently, the extended dynamics and cost functional of a representative minor bank are as in
\begin{align}
d X_t^{i,\delta_i} =& \begin{bmatrix}
d x_t^{i,\delta_i}\\
d X_t^{0,\delta_0}
\end{bmatrix}
=\left(\Tilde{A}X_t^{i,\delta_i}+\mb{B}u^i_t+\Tilde{B}\bar{u}_t+\delta_i\mb{B}\omega^i \right)d t + \Sigma d \overline{W}_t^i, \label{MinorExtendStateDynamicsPerturbed}\allowdisplaybreaks\\
J^i(u^i+\delta_i\omega^i) = \tfrac{1}{2}\mathbb{E} \bigg[\int_0^T &\Big \{(X_s^{i,\delta_i})^\intercal \mb{Q} X_s^{i,\delta_i} + 2(X_s^{i,\delta_i})^\intercal \mb{N} \big(u_s^i +\delta_i \omega^i_s \big) +\big(u^i_s + \delta_i \omega^i_s \big)^2\Big \} ds + \big(X_T^{i,\delta_i}\big)^\intercal \widehat{Q} X_T^{i,\delta_i}  \bigg], \label{MinorExtendCostPerturbed}
\end{align}
\begin{gather}
 \tilde{A} = \left[ \begin{array}{cc}
-a & [a G,a F] \\
0 &  \tilde{A}_0-\mb{B}_0\mb{N}_0^\intercal -\mb{B}_0\mb{B}_0^\intercal\phi_t^0\mb{B}_0^\intercal \tilde{A}_0
\end{array} \right]\!,
\quad
 \mb{B} = \left[ \begin{array}{c} 1 \\ {0} \end{array}\right]\!, 
\quad
 \tilde{B} = \left[ \begin{array}{c} 0 \\ \tilde{B}_0 \end{array}\right]\!, 
\quad
\Sigma = \left[ \begin{array}{cc}
\sigma & 0 \\
0 &  \Sigma_0
\end{array} \right],\label{MinorExtendDynmaicMatrices}\notag\\
 \mb{Q} = \left[ \begin{array}{ccc}
\epsilon & -G \epsilon & -F \epsilon\\
-G \epsilon & G^2 \epsilon & F G \epsilon\\
-F \epsilon & F G \epsilon & F^2 \epsilon 
\end{array} \right]\!,
\quad \mb{N}=\left[ \begin{array}{c} q \\ -q G \\ -q F \end{array}\right]
\!,
\quad
 \widehat{Q} = \left[ \begin{array}{ccc}
c & -c G & -c F \\
-c G & c G^2 & c F G\\
-c F & c F G & c F^2
\end{array} \right]\!,\quad \overline{W}^i_t = \left[\begin{array}{c} W^i_t \\ \overline{W}^0_t\end{array}\right].
\end{gather}

\noindent \underline{\textbf{Step (iv)}}: We use the results  developed by \citet[Theorem 2, 3, 4]{Firoozi2020ConvexanalysisLQG} to obtain the best response strategy for the minor bank. For the LQG system \eqref{MinorDynamicsOriginal_i} and \eqref{MinorExtendCostPerturbed} the G\^ateaux derivative is given by
\begin{equation}\label{MinorGatDeriv}
\langle \mcD{J_i^\infty(u)}, \omega^i \rangle
 = \mathbb{E} \bigg[ \int_0^T\!\! \omega_t^i \bigg\{\mb{N}^\intercal X_t^i+u_t^i+\mb{B}^\intercal \bigg(e^{-\tilde{A}^\intercal t}M_t^i-\int_0^{t} e^{\tilde{A}^\intercal(s-t)}(\mb{Q} X_s^i+\mb{N} u_s^i) ds \bigg) \bigg\} \bigg] dt,
\end{equation}
where $M_t^i$ is a martingale defined as 
\begin{equation}\label{MinorMartingale} 
   M_t^i =  \mathbb{E} \Big[e^{{\tilde{A}^\intercal}} \widehat{Q}  X_T^i +\int_0^T e^{\tilde{A} s}(\mb{Q} X_s^i+\mb{N} u_s^i) ds |\mc{F}_s \Big].
\end{equation}
By the martingale representation theorem, we have
\begin{equation}\label{martingale_rep}
    M_t^i = M_0^i+\int_0^t Z_s^i d\overline{W}_s^i.
\end{equation}
By setting the perturbation $\delta_i$ in \eqref{MinorExtendStateDynamicsPerturbed} to zero, we get the unperturbed extended state dynamics for minor banks as
\begin{equation}\label{MinorExtendStateDynamics}
d X_t^i =\left(\Tilde{A}X_t^i+\mb{B}u^i_t+\Tilde{B}\bar{u}_t\right)d t + \Sigma d \overline{W}_t^i.
\end{equation}
From Theorem 3 in \citep{Firoozi2020ConvexanalysisLQG}, the minor bank's optimal control action is given by
\begin{equation}\label{MinorOptimalControlExtend}
    u_t^{i,*}=-\left(\mb{N}^\intercal X_t^i+\mb{B}^\intercal \bigg( e^{-{\tilde{A}^\intercal}}M_t^i-\int_0^t e^{\tilde{A}^\intercal(s-t)}(\mb{Q} X_s^i+ \mb{N} u_s^{i,*} )d s \bigg) \right).
\end{equation}
Then we define the minor bank's adjoint process $p_t^i$ by 
\begin{equation}\label{MinorAdjointProcess}
    p_t^i=e^{-{\tilde{A}^\intercal}}M_t^i-\int_0^t e^{\tilde{A}^\intercal(s-t)}(\mb{Q} X_s^i+ \mb{N} u_s^{i,*} )ds.
\end{equation}
Next we adopt the ansatz 
\begin{equation}\label{MinorAnsatz}
    p_t^i = \Phi_t\left(\big(F \bar{x}_t + G x_t^0\big)-x_t^i\right),
\end{equation}
where $\Phi_t^\intercal = \begin{bmatrix} \phi_t & \psi_t & \lambda_t \end{bmatrix}$.
This ansatz can be equivalently represented as
\begin{equation}
p^i_t = -\frac{1}{q}\Phi_t \mb{N}^\intercal X_t^i.
\end{equation}
Subsequently, the optimal control action \eqref{MinorOptimalControlExtend} may be represented as 
\begin{equation}
\begin{aligned}
    u_t^{i,*} &= -\Big(\mb{N}^\intercal X_t^i+\mb{B}^\intercal p_t^i \Big)\\
        &= - \left(\mb{N}^\intercal X_t^i-\frac{1}{q}\mb{B}^\intercal \Phi_t \mb{N}^\intercal X_T^i \right)\\&=\big(q-\phi_t\big) \left( \big(F \bar{x}_t +G x_t^0\big)-x_t^i\right).\label{MinorOptimaControl}
        \end{aligned}
\end{equation}
The mean field $\bar{u}_t$ of the optimal control actions can then be computed by taking the conditional expectation of \eqref{MinorOptimaControl}   given $\mc{F}^0_t$. Let us define
$K^\intercal=[
        0 \,\, \,\,  G \,\, \,\,  F-1 ]$. Then, this calculation results in
\begin{equation}\label{MeanFiledOptimalControl}
\begin{split}
    \bar{u}_t&=(q-\phi_t)\left(G x_t^0+(F-1)\bar{x}_t\right)\\
    &=\frac{1}{a}\big(q-\phi_t\big)\tilde{B}_0^\intercal \tilde{A}_0 X_t^0,\\
    &=\big(q-\mb{B}^\intercal \Phi_t\big) K^\intercal X_t^i,
\end{split}
\end{equation}
We apply Ito's Lemma to \eqref{MinorAdjointProcess} and utilize \eqref{martingale_rep} to find the SDE that $p_t^i$ satisfies 
\begin{equation}\label{MinorSdeAdjointMedium}
dp_t^i=\left(-\tilde{A}^\intercal p_t^i -\left(\mathbb{Q}X_t^i+\mathbb{N}u_t^{i,*} \right) \right) d t + e^{-{\tilde{A}^\intercal}t}Z_t^i d\overline{W}_t^i.
\end{equation}
Substituting \eqref{MinorAnsatz} into \eqref{MinorSdeAdjointMedium} yields
\begin{equation}\label{MinorSdeAdjoint}
d p_t^i=\left(\frac{1}{q}\tilde{A}^\intercal \Phi_t \mathbb{N}^\intercal- \mathbb{Q} + \mathbb{N} \mathbb{N}^\intercal-\frac{1}{q} \mathbb{N} \mathbb{B}^\intercal \Phi_t \mathbb{N}^\intercal \right)X_T^i d t + e^{-{\tilde{A}^\intercal}t}Z_t^i d\overline{W}_t^i.
\end{equation}
Furthermore, applying Itô's Lemma to \eqref{MinorAnsatz} and the $p_t^i$ process leads to another SDE as in
\begin{equation}\label{MinorSdeAnsatzMedium}
\begin{aligned}
d p_t^i= \left(-\frac{1}{q} \dot{\Phi_t}\mathbb{N}^\intercal X_t^i-\frac{1}{q} \Phi_t \mathbb{N}^\intercal \left(\tilde{A} X_t^i +\mathbb{B}u_t^i+\tilde{B}\bar{u}_t \right) \right) d t-\frac{1}{q}\Phi_t \mathbb{N}^\intercal \Sigma d\overline{W}_t^i.
\end{aligned}
\end{equation}
Substituting \eqref{MinorOptimalControlExtend} and \eqref{MeanFiledOptimalControl} into the drift term of \eqref{MinorSdeAnsatzMedium} results in
\begin{equation}
\label{MinorSdeAnsatzFinal}
d p_t^i= \Big(-\frac{1}{q} \dot{\Phi_t}\mb{N}^\intercal-\frac{1}{q} \Phi_t \mb{N}^\intercal \tilde{A} +\frac{1}{q} \Phi_t \mb{N}^\intercal \mb{B}\mb{N}^\intercal - \frac{1}{q^2}\Phi_t \mb{N}^\intercal \mb{B}\mb{
B}^\intercal \Phi_t \mb{N}^\intercal-\frac{1}{q}\big(q-\mb{B}^\intercal \phi_t\big)\Phi_t \mb{N}^\intercal \tilde{B} K^\intercal \Big) X_t^i d t-\frac{1}{q}\Phi_t \mb{N}^\intercal \Sigma d\overline{W}_t^i. 
\end{equation}
We then match the two SDEs, \eqref{MinorSdeAdjoint} and \eqref{MinorSdeAnsatzFinal}, to obtain the conditions that $\Phi_t$ must satisfy

\begin{equation}\label{MinorConditionDiffusion}
 -\frac{1}{q}\Phi_t \mb{N}^\intercal \Sigma = e^{-{\tilde{A}^\intercal}t}Z_t^i.
\end{equation}\begin{equation}\label{MinorConditionDrift}
\begin{split}
-\frac{1}{q} \dot{\Phi_t}\mb{N}^\intercal=\frac{1}{q} \Phi_t \mb{N}^\intercal\tilde{A} -\frac{1}{q} \Phi_t \mb{N}^\intercal \mb{B} \mb{N}^\intercal + \frac{1}{q^2}\Phi_t \mb{N}^\intercal \mb{B}\mb{
B}^\intercal \Phi_t \mb{N}^\intercal+\frac{1}{q}\big(q-\mb{B}^\intercal \Phi_t\big)\Phi_t \mb{N}^\intercal \tilde{B} K^\intercal \\+\frac{1}{q}\tilde{A}^\intercal \Phi_t \mb{N}^\intercal- \mb{Q} + \mb{N} \mb{N}^\intercal-\frac{1}{q} \mb{N} \mb{B}^\intercal \Phi_t \mb{N}^\intercal. 
\end{split}
\end{equation}
which, respectively, result from equating the diffusion and drift coefficients.

In this section, we characterized the optimal control actions of minor banks, which are used to derive the mean-field of control actions $\bar{u}_t$. These results will be utilized in the next section to complete the solution of the major bank's problem.

\subsection{Return to Major Bank's Problem}

\vspace{0.2cm}
In this section, we recall some results from the previous sections and perform calculations for the major bank's LQG system. Then, we derive the optimal control for the major bank. First, we substitute $\bar{u}_t$ \eqref{MeanFiledOptimalControl} in unperturbed extended extended dynamics of the major given by \eqref{MajorExtendStateDynamicOriginal} to get 
\begin{equation} \label{MajorExtendStateDynamicFinal}
\begin{split} 
d X_t^0 &=\bigg( \Big(\Tilde{A}_0+\frac{1}{a}(q-\phi_t)\tilde{B}_0 \tilde{B}^\intercal_0 \tilde{A}_0 \Big) X_t^0+ \mb{B}_0u^0_t \bigg) d t+ \Sigma_0 d\overline{W}_t^0 \allowdisplaybreaks\\
&= \left( \mathbb{A}_0 X_t^0+ \mb{B}_0 u_t^0 \right) d t +\Sigma_0 d\overline{W}_t^0,
\end{split}
\end{equation}
where $\mb{A}_0=\Tilde{A_0}+\frac{1}{a}(q-\phi_t)\tilde{B}_0 \tilde{B}_0^\intercal \tilde{A}_0$. Moreover, we recall the cost functional from the previous section given by \eqref{MajorExtendCostPerturbed}.

\noindent \underline{\textbf{Step (i)}}: We calculate the G\^ateaux derivative of the major bank's cost functional as in
\begin{equation}\label{MajorGatDeriv}
\langle \mcD{J_0^\infty(u)}, \omega^0 \rangle
 = \mathbb{E} \bigg[ \int_0^T\!\! \omega_t^0 \bigg\{ \mb{N}^\intercal X_t^0+u_t^0+\mb{B}^\intercal \bigg(e^{-\mb{A}^\intercal t}M_t^0- \int_0^{t} e^{\mb{A}^\intercal(s-t)}(\mb{Q}_0 X_s^i+\mb{N}_0 u_s^i) ds \bigg) \bigg\} \bigg] dt,
\end{equation}
where $M_t^0$ is a martingale given by 
\begin{equation}\label{MajorMartingale} 
   M_t^0 =  \mathbb{E} \Big[e^{{\mb{A}_0^\intercal}} \mb{G}_0  X_T^0 +\int_0^T e^{\mb{A}_0^\intercal s}(\mb{Q}_0 X_s^0+\mb{N}_0 u_s^0) ds |\mc{F}_s \Big].
\end{equation}
By the martingale representation theorem we have
\begin{equation}\label{martingale_rep_major}
    M_t^0 = M_0^0+\int_0^t Z_s^0 d\overline{W}_s^0.
\end{equation}

\noindent \underline{\textbf{Step (ii)}}: 
From Theorem 3 in \citep{Firoozi2020ConvexanalysisLQG}, we obtain the major agent's optimal control action given by
\begin{equation}\label{MajorOptimalControlExtend}
    u_t^{0,*}=-\left(\mb{N}_0^\intercal X_t^0+\mb{B}_0^\intercal \bigg( e^{-{\mb{A}_0^\intercal}}M_t^0-\int_0^t e^{\mb{A}_0^\intercal(s-t)}(\mb{Q}_0 X_s^0+ \mb{N}_0 u_s^{0,*} )d s \bigg) \right).
\end{equation}
\noindent \underline{\textbf{Step (iii)}}: We aim to obtain  a linear State feedback representation for the major banks optimal control action. For this purpose, we define the major bank's adjoint process $p_t^0$  by
\begin{equation}\label{MajorAdjointProcess}
    p_t^0=e^{-{\mb{A}_0^\intercal}}M_t^0-\int_0^t e^{\mb{A}_0^\intercal(s-t)}(\mb{Q}_0 X_s^0+ \mb{N}_0 u_s^{0,*} )ds.
\end{equation}
We then adopt the ansatz
\begin{equation}\label{MajorAnsatz}
    p_t^0 =-\frac{1}{q_0}\Phi^0_t \mb{N}_0^\intercal X_t^0=\Phi_t^0\big(\bar{x}_t - x_t^0 \big),
\end{equation}
where $(\Phi_t^0)^\intercal=[ \phi_t^{0} \,\, \psi_t^{0}] 
$.
Substituting \eqref{MajorAdjointProcess} and \eqref{MajorAnsatz} into \eqref{MajorOptimalControlExtend} results in

\begin{equation}\label{MajorOptimalControlExtend_rep}
\begin{aligned}
    u_t^{0,*}&=-\big(\mb{N}_0^\intercal X_t^0+\mb{B}_0^\intercal p_t^0 \big)\allowdisplaybreaks
    \\&=-\big(\mb{N}_0^\intercal X_t^0+\mb{B}_0^\intercal \Phi_t^0\big( \bar{x}_t - x_t^0 \big) \big)\allowdisplaybreaks \\
    &=\big(q-\phi^0_t\big) \big(\bar{x}_t - x_t^0\big)\allowdisplaybreaks.
\end{aligned}
\end{equation}
We apply Ito's lemma to \eqref{MajorAdjointProcess} and use \eqref{martingale_rep_major} to get the SDE that $p_t^0$ satisfies as in
\begin{equation}\label{MajorAdjointSdeMedium}
    d p_t^0=\Big(-\mb{A}_0^\intercal p_t^0-\Big(\mb{Q}_0X_t^0+\mb{N}_0 u_t^{0,*} \Big) \Big) d t + e^{-\mb{A}_0^\intercal t} Z_t^0 d\overline{W}_t^0.
\end{equation}
Then we substitute \eqref{MajorAnsatz} and \eqref{MajorOptimalControlExtend} in  \eqref{MajorAdjointSdeMedium} to get
\begin{equation}\label{MajorSdeAdjoint}
    d p_t^0=\Big(-\frac{1}{a_0} \mb{A}_0^\intercal \Phi_t^0 \mb{B}_0^\intercal \tilde{A}_0 X_t^0 -\mb{Q}_0 X_t^0 +\mb{N}_0 \mb{N}_0^\intercal X_t^0 +\frac{1}{a_0}\mb{N}_0 \mb{B}^\intercal \Phi_t^0 \mb{B}_0^\intercal \tilde{A}_0 X_t^0 \Big) d t + e^{-\mb{A}_0^\intercal t} Z_t^0 d\overline{W}_t^0.
\end{equation}
Moreover, we apply Ito's Lemma to \eqref{MajorAnsatz} to obtain another SDE that $p_t^0$ satisfies as in
\begin{multline}\label{MajorAnsatzSdeMedium}
    d p_t^0=\Big(\dot{\Phi}_t^0 \big( \bar{x}_t - x_t^0 \big)+\Phi_t^0  \big( a(F-1) \bar{x}_t +a G x_t^0 +\bar{u}_t\big)-\Phi_t^0 \big(a_0  \bar{x}_t -a_0 x_t^0 +u_t^0 \big) \Big) d t -\phi_t^0\sigma_0  d\overline{W}_t^0.
\end{multline}
We then rewrite \eqref{MajorAnsatzSdeMedium} in terms of the extended state $X_t^0$ and substitute \eqref{MajorOptimalControlExtend_rep} and \eqref{MeanFiledOptimalControl} to get
\begin{equation}\label{MajorSdeAnsatz}
    d p_t^0=\bigg( \left(\frac{1}{a_0}\dot{\Phi}_t^0-\Phi_t^0 +\frac{1}{a_0}\Phi_t^0 \mb{B}^\intercal\phi_t^0 \right) \mb{B}_0^\intercal \tilde{A}_0 X_t^0 +\left(1 + \frac{1}{a} \big( q-\phi_t \big)\right) \Phi_t^0\tilde{B}_0^\intercal \tilde{A}_0 X_t^0 +\Phi_t^0 \mb{N}_0^\intercal X_t^0 \bigg) d t-\Phi_t^0\sigma_0 d\overline{W}_t^0.
\end{equation}
Finally, we match the SDEs \eqref{MajorSdeAdjoint} and \eqref{MajorSdeAnsatz} to get the two conditions that $\Phi_t^0$ must satisfy, i.e.
    \begin{align}
    &\hspace{2cm} e^{-\mb{A}_0^\intercal t} Z_t^0=-\Phi_t^0\sigma_0 \label{MajorConditionDiffusion}\\
    \big(\frac{1}{a_0}\dot{\Phi}_t^0-\Phi_t^0 &+\frac{1}{a_0}\Phi_t^0 \mb{B}^\intercal\Phi_t^0 \big) \mb{B}_0^\intercal \tilde{A}_0 +\Phi_t^0 \tilde{B}_0^\intercal\tilde{A}_0  + \frac{1}{a} \big(q-\phi_t \big) \Phi_t^0 \tilde{B}_0^\intercal \tilde{A}_0  +\Phi_t^0 \mb{N}_0^\intercal  \notag\\&\hspace{3cm}=-\frac{1}{a_0} \mb{A}_0^\intercal \Phi_t^0 \mb{B}_0^\intercal \tilde{A}_0  -\mb{Q}_0  +\mb{N}_0 \mb{N}_0^\intercal  +\frac{1}{a_0}\mb{N}_0 \mb{B}^\intercal \Phi_t^0 \mb{B}_0^\intercal \tilde{A}_0.\label{MajorConditionDrift}
    \end{align}

To conclude, we derived the optimal trading strategies for the major bank and a representative small bank $\mc{A}_i$ given, respectively, by \eqref{MajorOptimalControlExtend_rep}, \eqref{MajorConditionDrift}, and \eqref{MinorOptimaControl}, \eqref{MinorConditionDrift}. We can then exploit the structure of system matrices to simplify the optimal strategies and the associated ODEs through matrix multiplications. This leads to a reduced representation of the optimal trading strategies as given in the statement of Theorem \ref{thm:Best_resp}. 

\section{Additional Numerical Experiments}

In this appendix, we report results of additional numerical experiments. frequency of trading (mean-reversion rate), it tends to converge more quickly to the market state brought down by a defaulting major bank. Hence, this negative externality offsets the benefit the minor bank obtains from having a major bank in the market. Therefore, the total default probability is higher than that of a market without a major bank for each value of $a$, and it decreases with $a$ with a smaller slope.

\subsection{Quality of Approximation}\label{app:approxquality}
We first examine the convergence of the simulation results obtained for the finite and infinite population cases. For the finite population, as described by \eqref{MajorDynamicsOriginal}-\eqref{MinorCostOriginal} in \cref{section:Model Description}, we consider a setup where there are $10$ minor banks and one major bank in the financial system and perform 50000 simulations for various settings. For the infinite population, as described by \eqref{major_dyn_inf}-\eqref{MeanFiledUbar} in \cref{section:MFG formulation}, we simulate $10^4$ minor banks in the economy and perform 5000 simulations for each market setting.

\begin{figure*}
    \centering
    \includegraphics[scale=0.5]{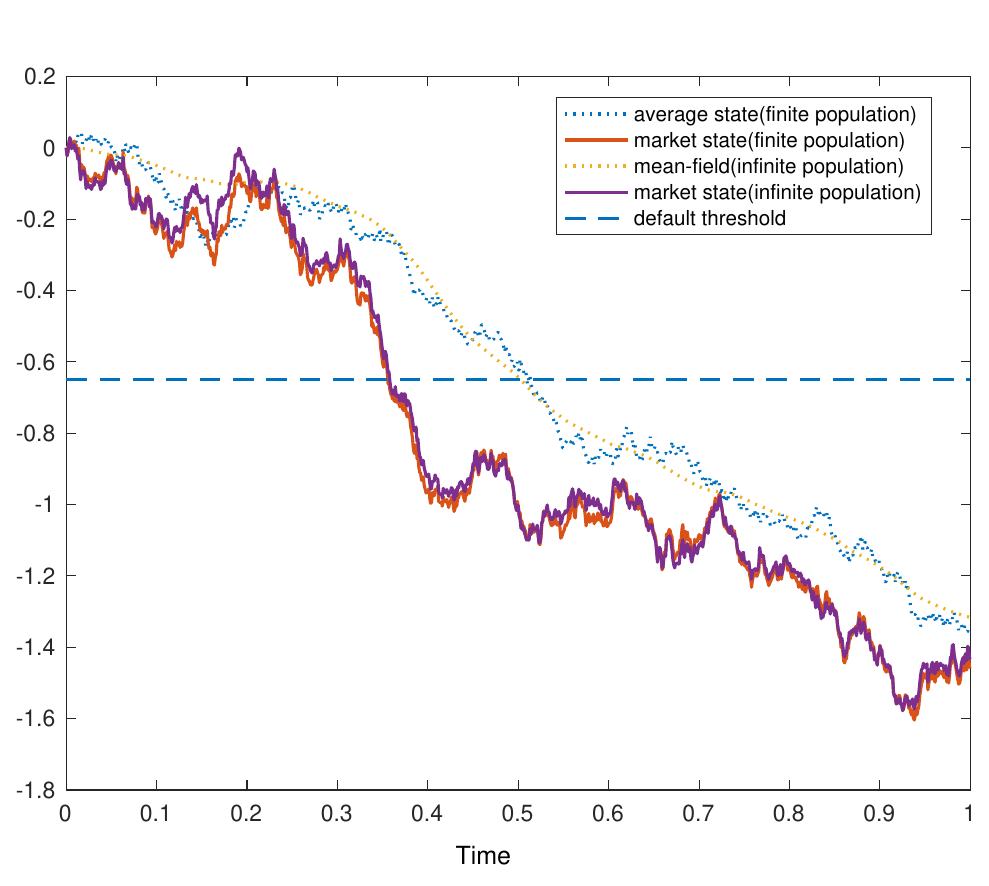}
    \caption{Convergence of the average state and the market state in the finite population to corresponding quantities in the infinite population.}
    \label{fig:MarketState-MeanField_infinite_finite}
\end{figure*}

Remark that the strategies employed by banks in the finite population correspond to the limiting strategies, where the mean-field $\bar{x}_t$ and the limiting log-monetary reserves are, respectively, replaced by the empirical average $x^{(N)}_t$ of the log-monetary reserves of minor banks and the log-monetary reserves specific to  the finite-population. 
Hence, the equilibrium of a large population game is here approximated by the Nash equilibrium of the limiting game when the number of agents goes to infinity. To investigate the quality of this approximation, 
we depict sample trajectories of the average state of minor banks and the market state, which is a linear combination of the average state of minor banks and the state of the major bank, in both finite and infinite population cases \cref{fig:MarketState-MeanField_infinite_finite}. We observe that the trajectories of the mass of small banks $(x^{(N)}_t,\bar{x}_t)$ and those of the market state $(F \bar{x}_t + G x_t^0,F {x}^{(N)}_t + G x_t^0 )$ evolve closely. Therefore, \cref{fig:MarketState-MeanField_infinite_finite} illustrates that the behavior of the system in the infinite population case is a good approximation to that in the finite population even when the number of minor banks in the finite population is relatively small.

In the remainder of this section, we conduct simulations for both the finite population and infinite population scenarios across various settings. The objective is to estimate the default probabilities as outlined in Section \ref{section:Individual Default and Systemic Risk in Interbank Transactions}, utilizing Monte Carlo simulations. Additionally, we explore the influence of the relative size $G$ and the mean reversion rate $a_0$ of the major bank.

\subsection{Role of the Major Bank}
In our model, 
each minor bank tracks the average log-monetary reserve in the market (or the market state) given by $(F {x}_t^{(N)} +G x_t^0)$ in the finite population and by $(F \bar{x}_t +G x_t^0)$ in the infinite population, respectively. This market state is a weighted average of the major bank's log-monetary reserve and the average level of log-monetary reserves across small banks. The parameters $G$ and $F$ denote, respectively, the relative sizes of major bank and the mass of small banks such that $G+F=1$. We perform simulations of the model for different values of $G \in \{0.1, 0.2, 0.3,\dots, 0.9\}$.

\subsubsection{Default Probability of a SmallMinor Bank}\label{app:defauly proba of a small bank}

We study the default probability of a representative small bank for different values of $G\in \{0.1, 0.2, \dots, 0.9\}$. The obtained results for the finite-population case are summarized in \cref{app:RegressionEstimate_FG_Finite:PDi}.

The average default probability of a representative minor bank is around 0.3405 in the absence of a major bank. This probability $\bar{p}_i$ increases when there exists a major bank in the market, except for the case where the major bank is relatively small with respect to the mass of small banks. Moreover, the probability of default increases further as the relative size $G$ of the major bank enlarges. Unconditionally, a major bank does not seem to improve the stability of small banks except if its size remains very limited.

Examining the default probabilities of the representative small bank in relation to whether the major bank has defaulted or not, denoted as $\bar{p}_{i|MD}$ and $\bar{p}_{i|MS}$ respectively, provides valuable insights. It shows that the default of a major bank significantly increases the likelihood of default for a small bank. The default probabilities range from 0.5112 to 0.7923, with a higher impact observed when the major bank is larger in size. This is due to the strong connections between a large major bank and smaller banks, leading to a higher level of exposure in the event of default.
Conversely, when a major bank remains stable, the default probability of a representative small bank improves from 0.3405 to approximately 0.20. This positive impact remains consistent across all relative sizes considered for the major bank.

From these initial simulations, we conclude that a major bank has two opposing effects on the default probability of a representative small bank. On the one hand, a successful (or stable) major bank improves slightly the position of the small bank, as it can provide additional liquidity when the small bank needs money to cover a liquidity shortage, and offers a coordination channel for small banks. On the other hand, the substantial negative externality that exists from the possible failure of a major bank puts the small bank at risk. 
In our setting, the net impact is negative as soon as the large bank represents more than 10\% of the  interbank market. The results for the infinite population are similar. 

\begin{table*}[h]
\centering
\scalebox{0.9}{
\begin{tabular}{ccccc}
\hline
 & No Major Bank  & \multicolumn{3}{c}{With a Major Bank}                                                                     \\ \cline{3-5} 
                
                   &                                & \multicolumn{1}{c}{}                       & \multicolumn{1}{c}{Non-defaulting Major} & Defaulting Major \\
                 G  &                     \multicolumn{1}{c}{$\bar{p}_i$}           & \multicolumn{1}{c}{$\bar{p}_i$}        & \multicolumn{1}{c}{$\bar{p}_{i|MS}$}     & $\bar{p}_{i|MD}$ \\ \hline
0                                        & 0.3388                         & \multicolumn{1}{c}{-}                              & \multicolumn{1}{c}{-}                    & -                                     \\ 
0.1                                      & -                              & \multicolumn{1}{c}{0.3397}                         & \multicolumn{1}{c}{0.2263}               & 0.4867                                \\ 
0.2                                      & -                              & \multicolumn{1}{c}{0.3527}                         & \multicolumn{1}{c}{0.1998}               & 0.5581                                \\ 
0.5                                      & -                              & \multicolumn{1}{c}{0.3916}                         & \multicolumn{1}{c}{0.1786}               & 0.7107                                \\ 
0.7                                      & -                              & \multicolumn{1}{c}{0.4060}                         & \multicolumn{1}{c}{0.1816}               & 0.7614                                \\ 
0.9                                      & -                              & \multicolumn{1}{c}{0.4183}                         & \multicolumn{1}{c}{0.1887}               & 0.7958                                \\ 
\hline
\end{tabular}}
\caption{Estimated default probability of a representative minor bank in the finite-population model for the cases (from left to right) with (i) no major bank, (ii) total default probability ($\bar{p}_{i}$) with a major bank, (iii) conditional default probability ($\bar{p}_{i|MS}$) with a non-defaulting major bank, and (iv) conditional default probability ($\bar{p}_{i|MD}$) with a defaulting major bank.}
\label{app:RegressionEstimate_FG_Finite:PDi}
\end{table*}

\subsubsection{Trajectories of Banks}

\begin{figure*}[h]
    \centering
    \subfigure[]{
    \begin{minipage}[]{.45\linewidth}
    \centering\includegraphics[scale=0.4]{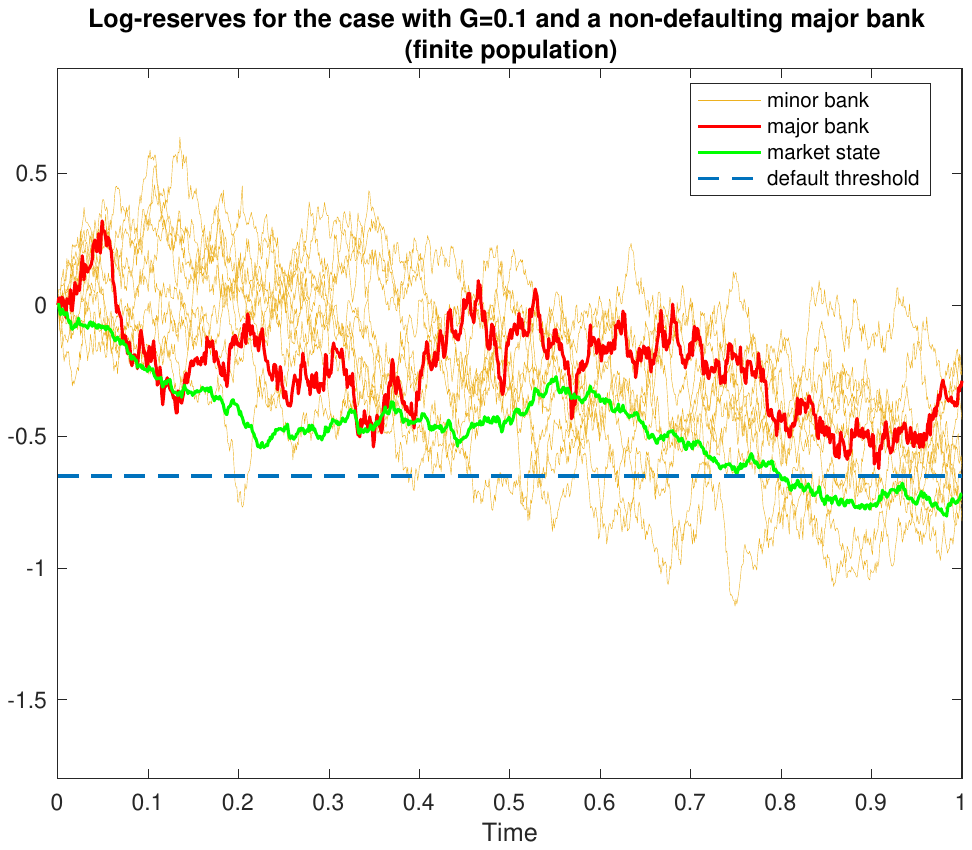}
    \end{minipage}
    }
    \subfigure[]{
    \begin{minipage}[]{.45\linewidth}
    \centering\includegraphics[scale=0.4]{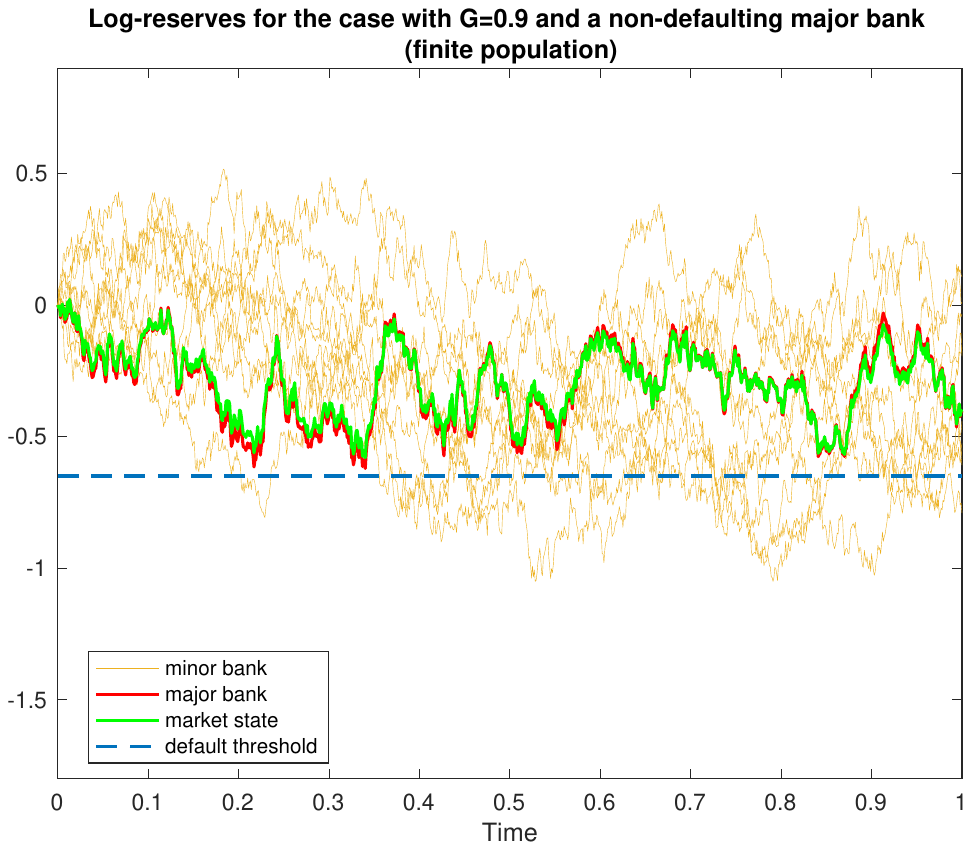}
    \end{minipage}
    }
    \subfigure[]{
    \begin{minipage}[]{.45\linewidth}
    \centering\includegraphics[scale=0.4]{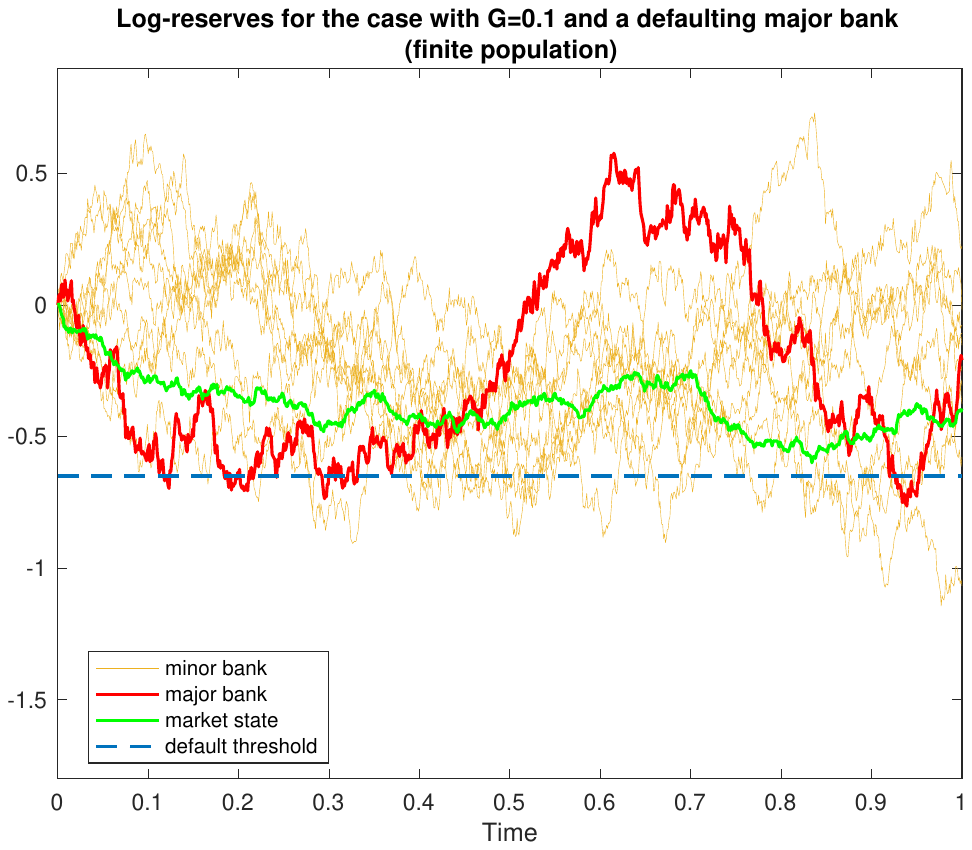}
    \end{minipage}
    }
    \subfigure[]{
    \begin{minipage}[]{.45\linewidth}
    \centering\includegraphics[scale=0.4]{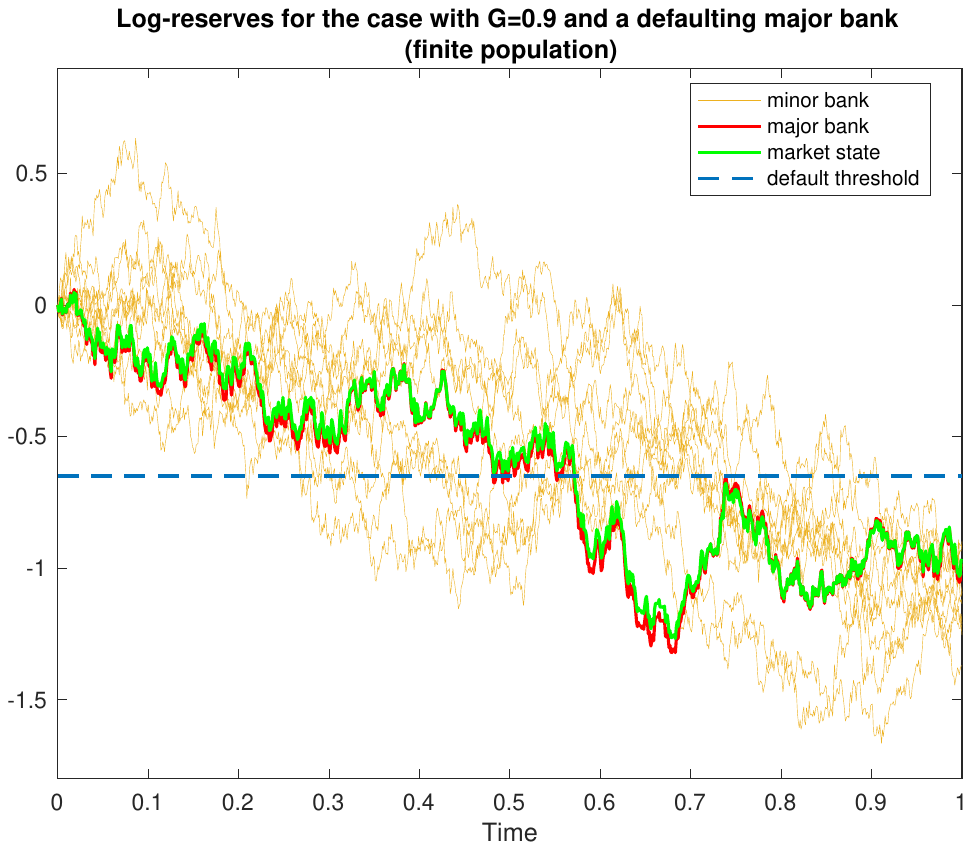}
    \end{minipage}
    }
    \caption{Simulated trajectories for 10 small banks, the major bank and the market state in the finite-population case with ($a=5$): (a) $G=0.1$ and a non-defaulting major bank, (b) $G=0.9$ and a non-defaulting major bank, (c) $G=0.1$ and a defaulting major bank, and (d) $G=0.9$ and a defaulting major bank. In all cases.}
    \label{fig:Infinite_Trajectoires_G}
\end{figure*}

We illustrate simulated trajectories of log-monetary reserves for $10$ small banks, the major agent, and the market state for one simulation in \cref{fig:Infinite_Trajectoires_G} for the finite-population setting. Throughout these simulations the realizations of the stochastic processes that model uncertainty are the same across considered scenarios. As the relative size of the major bank increases, all trajectories evolve more closely together. Furthermore, the larger the major bank the faster it brings the system towards a systemic default. This feature is not clear for smaller sizes such as $G=0.1$. 
These plots are in line with the discussion of the previous tables. 

\subsection{Role of Market Frictions}
In this section, we assume that the major bank and the mass of minor banks are of the same size ($F=G=0.5$) and examine the impact of reducing market frictions, by increasing the mean-reversion rate $a$. Recall that we interpret $a_0$ and $a$ as inverse measures of market frictions. A smaller value of $a$ means larger market friction for all banks, since $a_0$ and $a$ parameters are jointly determined with $F$ and $G$, as already established when discussing the market clearing condition \eqref{ClearingCondition1}-\eqref{ClearingCondition2}. We therefore interpret larger values for $a_0$ and $a$ as indicative of the degree to which banks rely on the interbank market to satisfy their financial obligations. In this case, we have $a_0=0.5a$. A higher mean-reversion rate translates into a higher frequency of lending and borrowing activities. Thus, the major bank trades at a lower frequency than a representative minor bank given the same distance from their respective tracking signal, respectively, $\bar{x}_t$ and $0.5(x^0_t+\bar{x}_t)$. This could be due to some market frictions and conditions  as explained in \cref{section:Market Clearing Condition}. To investigate the impact of the mean reversion rates on the system we consider $a \in \{1,2,\dots,10\}$.

\subsubsection{Default Probability of a Minor Bank}
In \cref{RegressionEstimate_a_Finite:PDi}, we present the simulated probabilities. The first and second columns demonstrate that the default probability of a minor bank decreases as the mean-reversion rate $a$ increases in the market, regardless of the presence of a major bank. This observation indicates that when a minor bank engages in more trading activities, it becomes less prone to individual default since it can respond more rapidly to negative shocks. However, we also observe that the impact of increasing $a$ on the unconditional probability of default is more pronounced in the absence of the major bank. These results can be explained by examining the conditional default probabilities. When the major bank does not default, the role played by $a$ is symmetric compared to the scenario without a major bank. Specifically, increasing $a$ from $1$ to $10$ leads to a decrease in the probability of default by approximately 20 percentage points. However, if the major bank defaults, the opposite effect occurs: the probability of default increases with $a$. This is because the log-reserve of the major bank is a crucial component of the market state, which the minor bank explicitly targets. Consequently, when a minor bank increases its trading rate $a$, it converges more rapidly to the market state, which is adversely affected by the default of the major bank. This negative externality offsets the benefit that the minor bank derives from the presence of a major bank in the market. As a result, the overall probability of default is higher compared to the scenario without a major bank, and its decrease with increasing $a$ is less pronounced. These findings hold true in the case of an infinite population as well.

\begin{table*}[h]
\centering
\scalebox{0.9}{
\begin{tabular}{ccccc}
\hline
 & No Major Bank  & \multicolumn{3}{c}{With a Major Bank}                                                                     \\ \cline{3-5} 
                
                   &                                & \multicolumn{1}{c}{}                       & \multicolumn{1}{c}{Non-defaulting Major} & Defaulting Major \\
                 $a$  &                     \multicolumn{1}{c}{$\bar{p}_i$}           & \multicolumn{1}{c}{$\bar{p}_i$}        & \multicolumn{1}{c}{$\bar{p}_{i|MS}$}     & $\bar{p}_{i|MD}$ \\ \hline
1                                        & 0.4312                         & \multicolumn{1}{c}{0.4480}                         & \multicolumn{1}{c}{0.2912}               & 0.6447                                \\ 
3                                        & 0.3896                         & \multicolumn{1}{c}{0.4182}                         & \multicolumn{1}{c}{0.2278}               & 0.6776                                \\ 
5                                        & 0.3413                         & \multicolumn{1}{c}{0.3931}                         & \multicolumn{1}{c}{0.1781}               & 0.7115                                \\ 
7                                        & 0.3008                         & \multicolumn{1}{c}{0.3703}                         & \multicolumn{1}{c}{0.1440}               & 0.7317                                \\ 
10                                       & 0.2505                         & \multicolumn{1}{c}{0.3474}                         & \multicolumn{1}{c}{0.1086}               & 0.7611                                \\ \hline
\end{tabular}}
\caption{Estimated default  probability of a representative minor bank in the finite-population model for the cases (from left to right) with (i) no major bank, (ii) total default probability ($\bar{p}_{i}$) with a major bank, (iii) conditional default probability ($\bar{p}_{i|MS}$) with a non-defaulting major bank, and (iv) conditional default probability ($\bar{p}_{i|MD}$) with a defaulting major bank.}
\label{RegressionEstimate_a_Finite:PDi}
\end{table*}

\begin{figure*}
    \centering
    \subfigure[]{
    \begin{minipage}[]{.45\linewidth}
    \centering\includegraphics[scale=0.4]{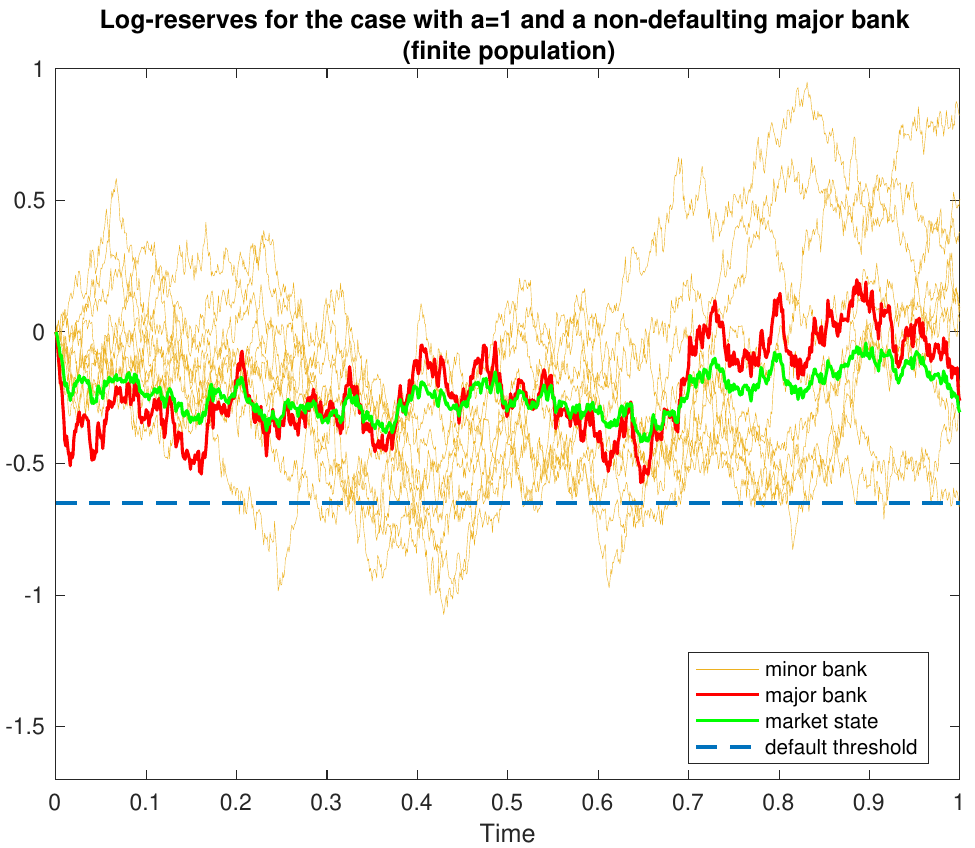}
    \end{minipage}
    }
    \subfigure[]{
    \begin{minipage}[]{.45\linewidth}
    \centering\includegraphics[scale=0.4]{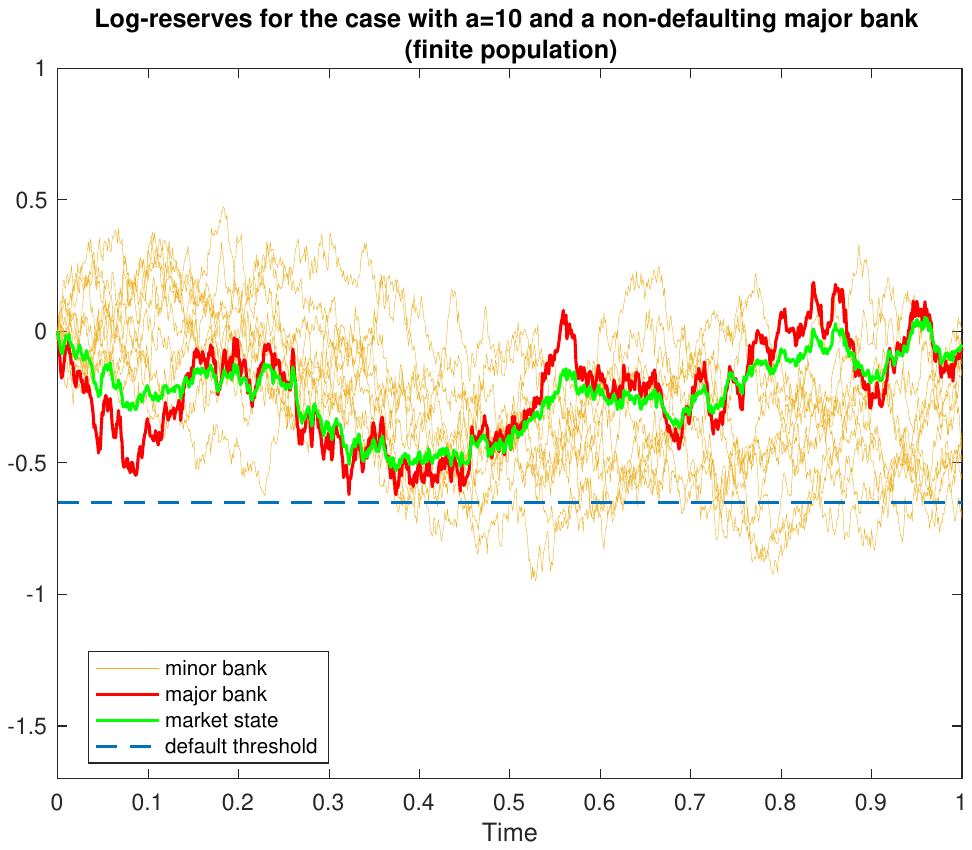}
    \end{minipage}
    }
    \subfigure[]{
    \begin{minipage}[]{.45\linewidth}
    \centering\includegraphics[scale=0.4]{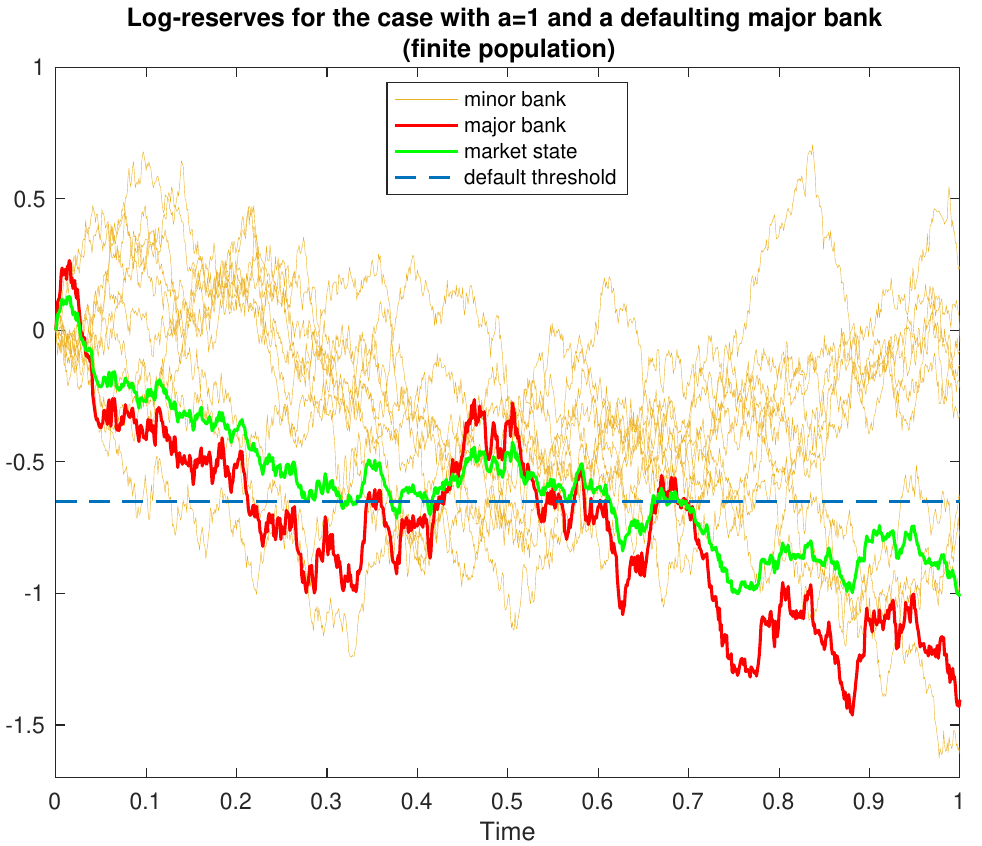}
    \end{minipage}
    }
    \subfigure[]{
    \begin{minipage}[]{.45\linewidth}
    \centering\includegraphics[scale=0.4]{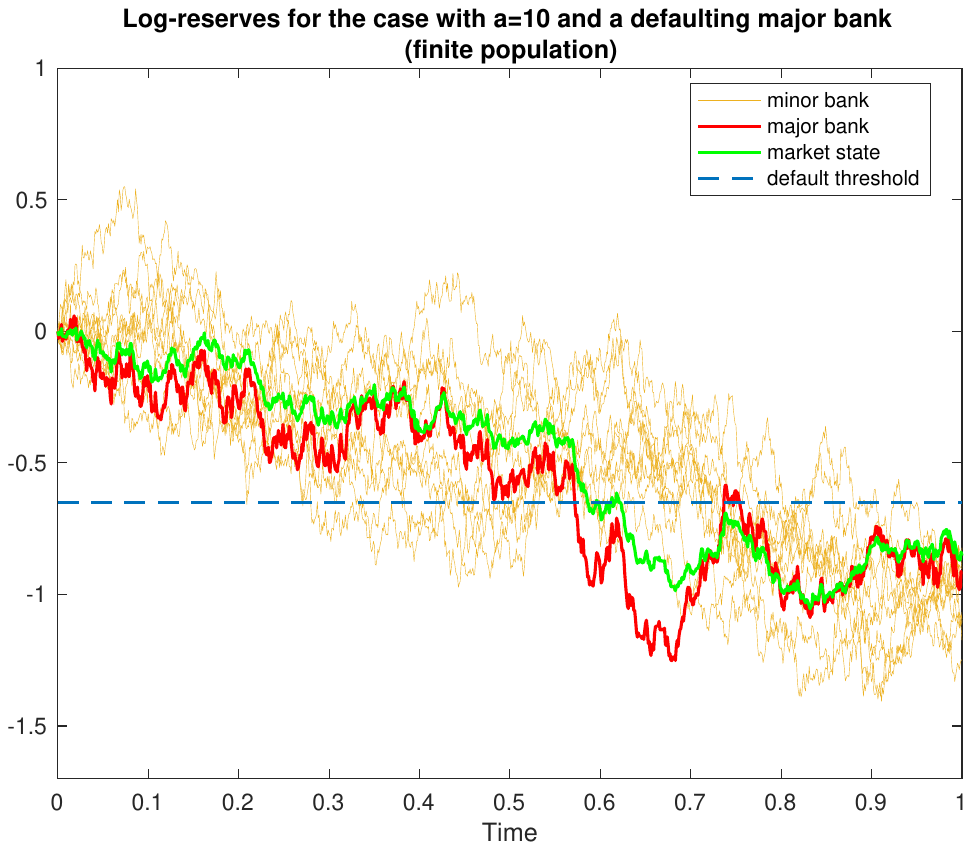}
    \end{minipage}
    }
    \caption{Simulated Trajectories for 10 small banks, the major bank and the market state in the finite-population case with ($G=0.5, F=0.5$): (a) $a=1$ and a non-defaulting major bank, (b) $a=10$ and a non-defaulting major bank, (c) $a=1$ and a defaulting major bank, and (d) $a=10$ and a defaulting major bank.}
    \label{fig:Infinite_Trajectoires_a}
\end{figure*}

\begin{figure*}[h]
    \centering
    \subfigure[$\phi_t \ and \ \phi_t^{0}$]{
    \begin{minipage}[]{.45\linewidth}
    \centering\includegraphics[scale=0.4]{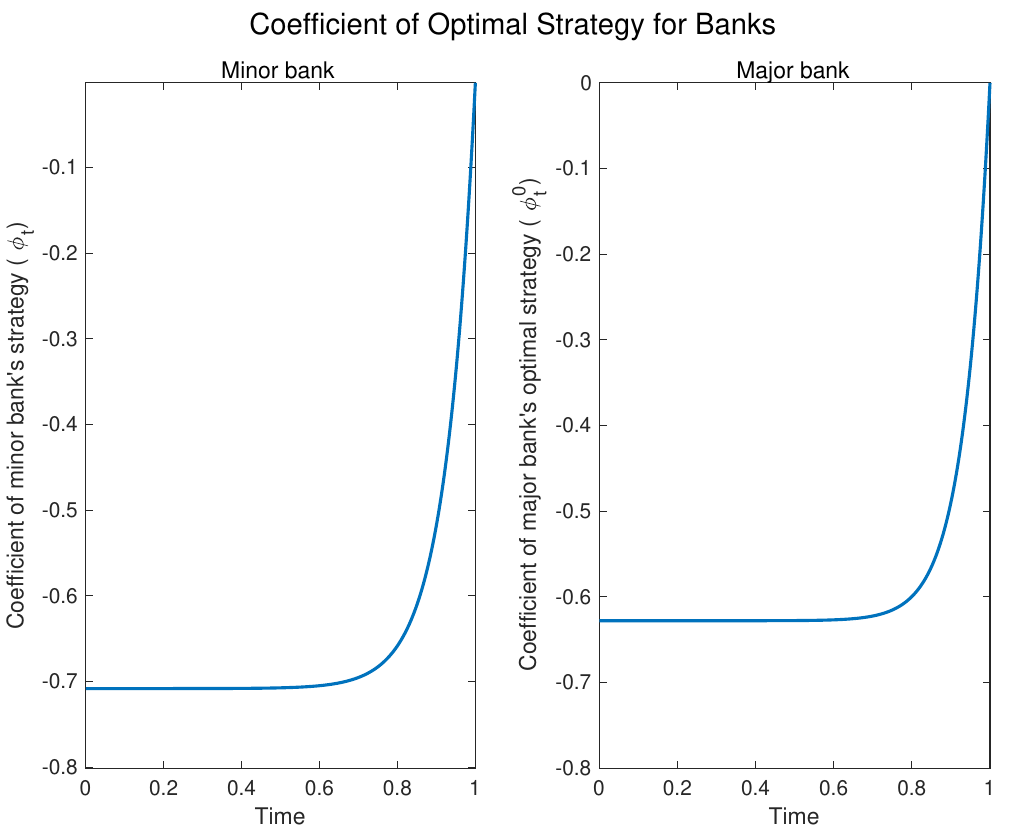}
    \end{minipage}
    }
    \subfigure[$\big(a+q-\phi_t \big) \ and \  \big(a_0 +q_0 -\phi^{0}_t \big)$]{
    \begin{minipage}[]{.45\linewidth}
    \centering\includegraphics[scale=0.4]{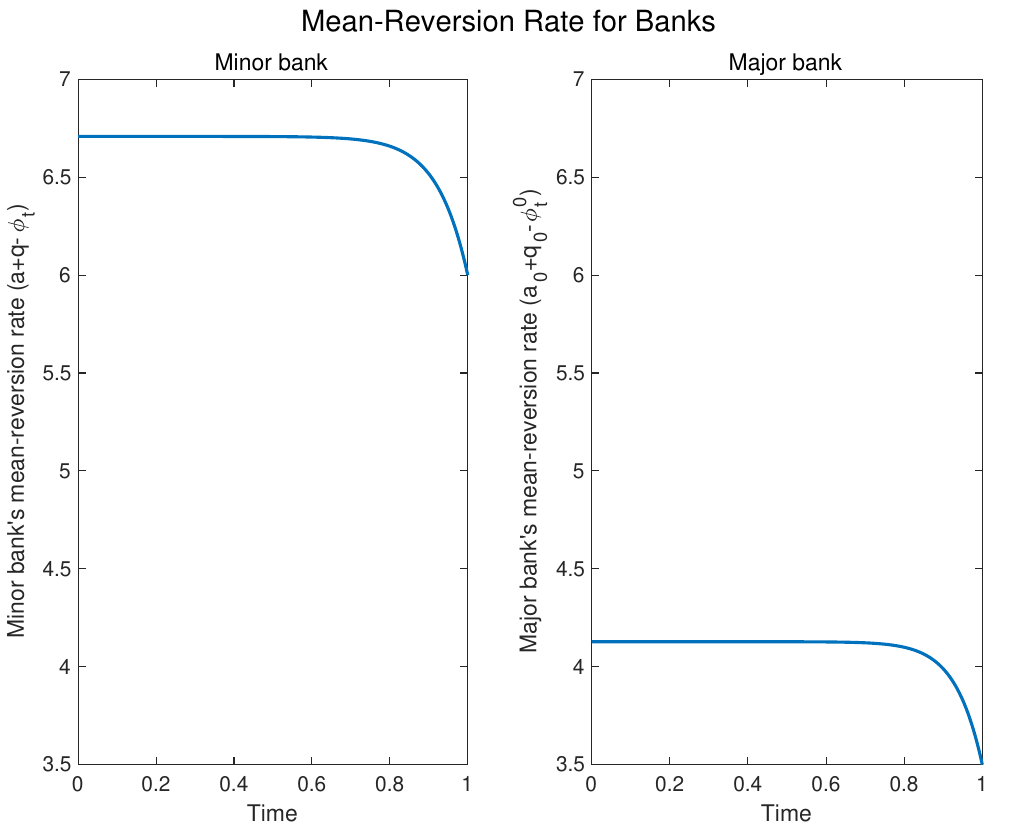}
    \end{minipage}
    }
    \caption{Simulation Results (from left to right): (a) solution of the ODE system, and (b) the mean-reversion level of different banks after adding the optimal control ($a=5,a_0=2.5,F=0.5,G=0.5,q=q_0=1$).} 
    \label{fig:phi-mean-reversion-level}
\end{figure*}

\subsubsection{Trajectories of Banks}
We illustrate the simulated trajectories for the log-monetary reserves of $10$ minor banks, the major bank, and the market state for the cases with $a=1$ and $a=10$ in \cref{fig:Infinite_Trajectoires_a}, for one simulation. Our results are  consistent with those in \citep{Fouque2013SystemicRiskIllustrated}. We find that there is a larger flocking effect such that the trajectories of minor banks evolve much more closely to each other as the mean-reversion rate $a$ increases from $1$ to $10$.  
Moreover, panels (c) and (d) reveal that a higher mean-reversion rate may delay the default of the major bank. However, when the major bank goes bankrupt it drags down the market state and hence the minor banks default much faster. This interpretation is possible because the realizations of stochastic processes modeling uncertainty are the same in both panels. 

\subsubsection{Transactions with the Central Bank}

We now investigate how optimal transactions with the central bank affect the log-monetary reserves. We find that the optimal strategies increase the mean-reversion rate by adding a time-varying component $\big( q_0-\phi_t^0 \big)$ and $\big( q-\phi_t \big)$, respectively, for the major bank and a representative minor bank. The evolution of $\phi_t^0$ and $\phi_t$ over time is depicted in panel (a) of \cref{fig:phi-mean-reversion-level}. Moreover, the evolution of the total mean reversion rates $\big(a+ q-\phi_t \big)$ and $\big( a_0 + q_0-\phi_t^0 \big)$ is depicted in panel (b) of \cref{fig:phi-mean-reversion-level}. We observe that the presence of a central bank provides the market participants with extra liquidity and increases the frequency of their transaction activities (note that $q-\phi_t>0$ and $q_0-\phi_t^0>0$). In our model, banks only trade during a fixed time period $[0,T]$ and they are not concerned about what happens after $T$. Banks start borrowing and lending activities with a higher mean-reversion rate since their long-term forecast is relatively imprecise. They prefer to trade at higher rates to compensate for the idiosyncratic shocks than to carry over these shocks until the end of the time period. However, as market uncertainty decreases towards the end of the trading period,  all banks naturally trade at smaller rates because idiosyncratic shocks have smaller dynamic consequences.

\begin{figure*}
    \centering
    \includegraphics[scale=0.5]{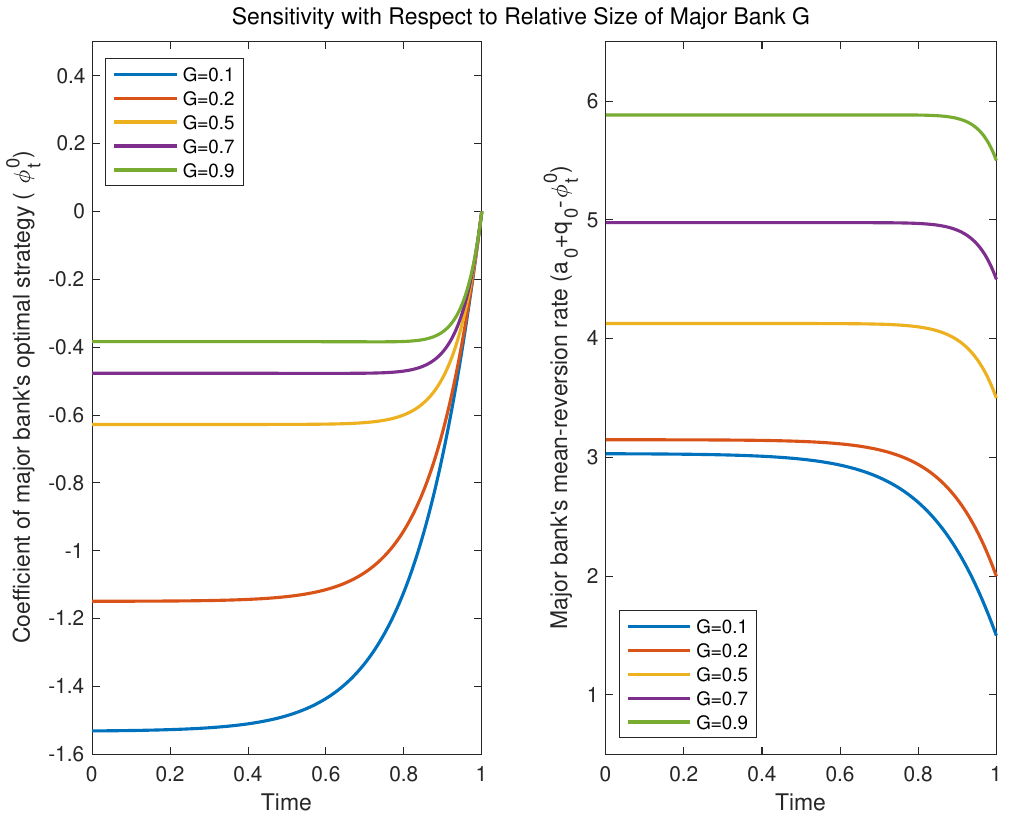}
    \caption{Sensitivity of the major agent's control coefficient $\phi^0_t$ and  mean-reversion rate $\big(a_0 +q_0 -\phi^{0}_t \big)$,  with respect to its relative size $G$, ($a=5, a_0=a*G, q_0=1, q=1, \epsilon_0=10, c_0=0$).}
    \label{fig:sensitivity_wrt_G}
\end{figure*}

 In Figure \ref{fig:sensitivity_wrt_G}, we illustrate how the mean-reversion rate of the major bank is influenced by the relative size of the major bank. The primary focus is on the evolution of $\phi_t^0$ over time given by \eqref{MajorRaccati_thm}, which begins at a negative value and gradually approaches zero. While the relative sizes of the banks, $F$ and $G$, do not alter the shape of $\phi_t^0$, they do impact the initial values. Specifically, as $G$ (the size of the major bank) increases, $\phi_t^0$ starts from a higher value, indicating a faster mean-reversion rate, defined by $a_0 + q_0 - \phi_t^0$.

More specifically, as seen in the right panel of Figure 6,  the expression $a_0 + q_0 - \phi_0$ increases with $G$. However, the left panel suggests that $q_0 - \phi_0$ decreases as $G$ grows. This apparent discrepancy arises because $a_0$ is not fixed; it varies with both $F$ and $G$, while $a$ is held constant because of the market-clearing condition given by \eqref{ClearingCondition1}-\eqref{ClearingCondition2}. Consequently, the figures might seem confusing, as they capture the interplay of multiple factors influencing the major bank's mean-reversion rate.

Consequently, the increase in the mean-reversion rate with $G$ is primarily driven by changes in $a_0$, which adjust based on the market-clearing conditions \eqref{ClearingCondition1}-\eqref{ClearingCondition2}. Although trades with the central bank, occurring at the rate $q_0-\phi^0_t$, act as a countervailing force, dampening the increase in mean-reversion rates, the effect of $a_0$ prevails, resulting in a net rise as $G$ grows. As the major bank expands, it engages less frequently with the central bank and it becomes a focal point for smaller banks, offering a stable market signal that facilitates coordination. This shift reduces the central bank’s direct role in market stabilization, allowing the interbank market to assume a greater relative size. Consequently, when the major bank becomes excessively large, the observed increase in systemic risk, as seen in \Cref{RegressionEstimate_FG_Finite:SystemicRisk}, reflects a scenario where the central bank is less actively involved, and the stability of the system increasingly depends on the actions of the major bank.

\subsubsection{Loss Distribution}
\begin{figure*}
    \centering
    \subfigure[loss distribution where there is no major bank]{
    \begin{minipage}[]{.45\linewidth}
    \centering\includegraphics[scale=0.395]{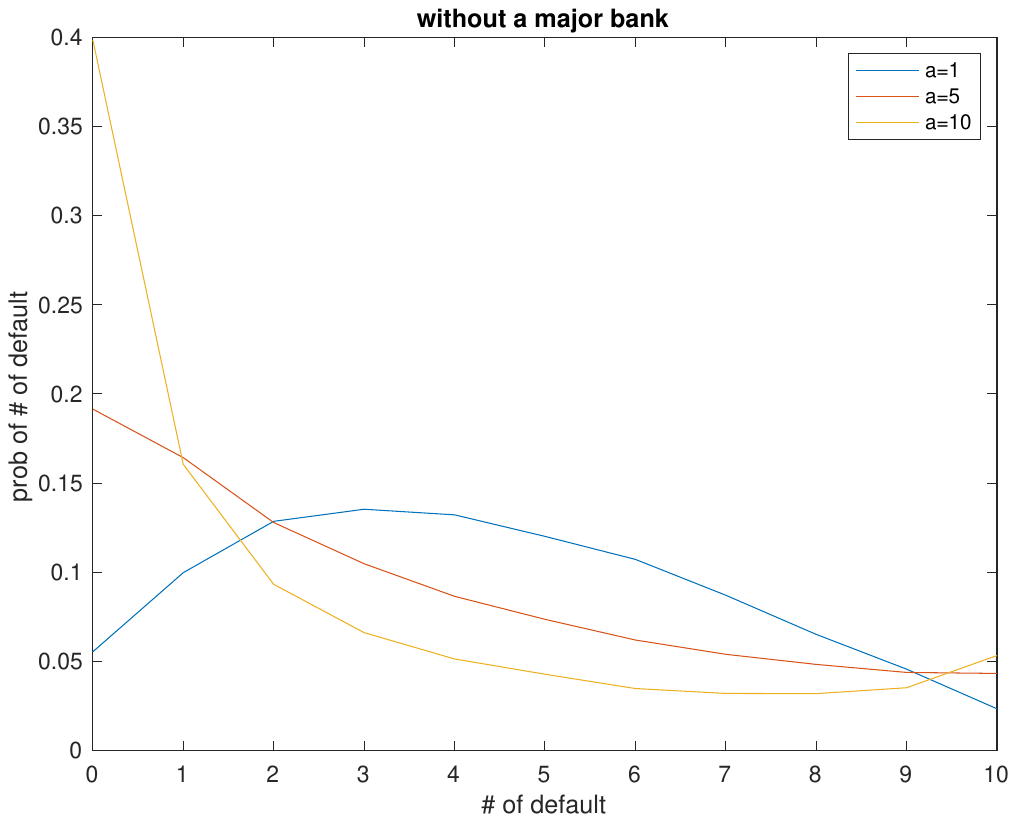}
    \end{minipage}
    }
     \subfigure[total loss distribution where there is a major bank ]{
    \begin{minipage}[]{.45\linewidth}
    \centering\includegraphics[scale=0.395]{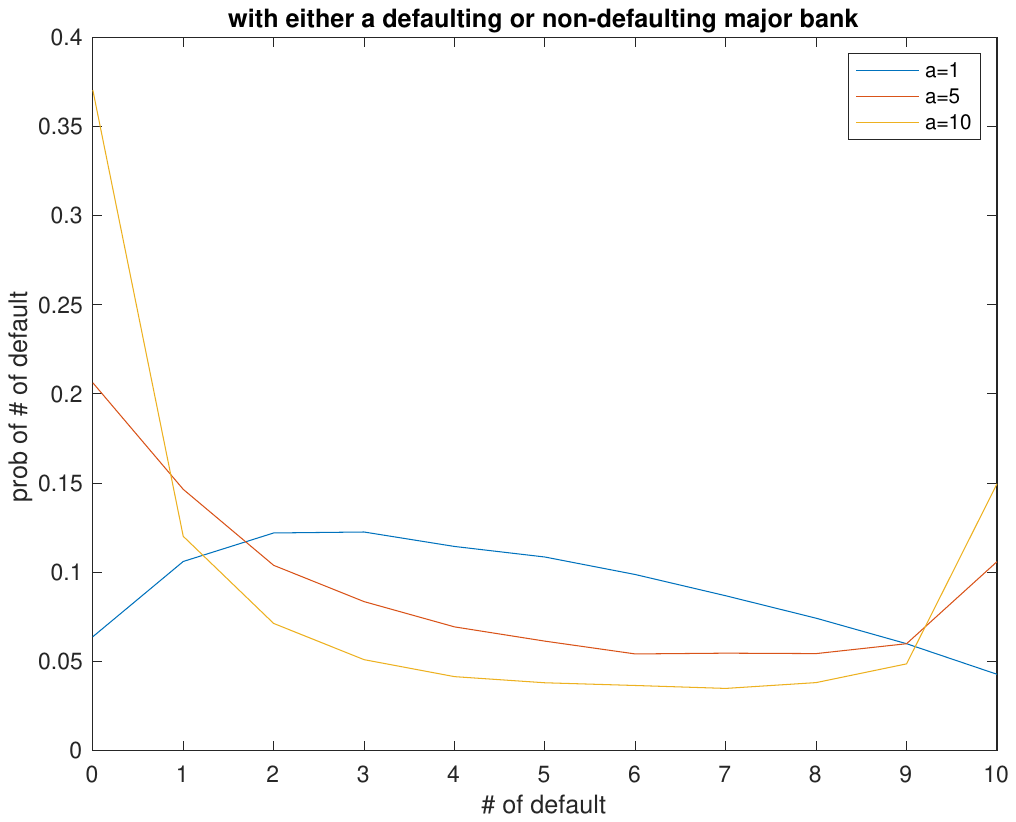}
    \end{minipage}
    }
    \subfigure[conditional loss distribution given that the major bank defaults]{
    \begin{minipage}[]{.45\linewidth}
    \centering\includegraphics[scale=0.395]{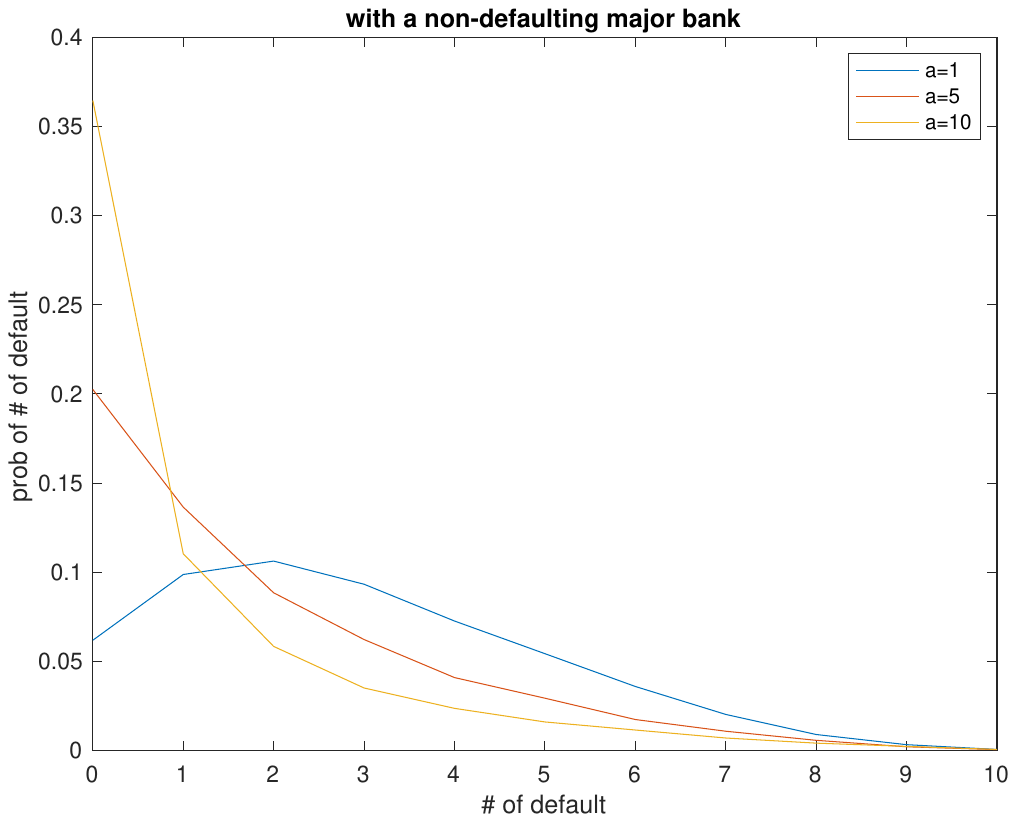}
    \end{minipage}
    }
    \subfigure[conditional loss distribution given that the major bank defaults]{
    \begin{minipage}[]{.45\linewidth}
    \centering\includegraphics[scale=0.395]{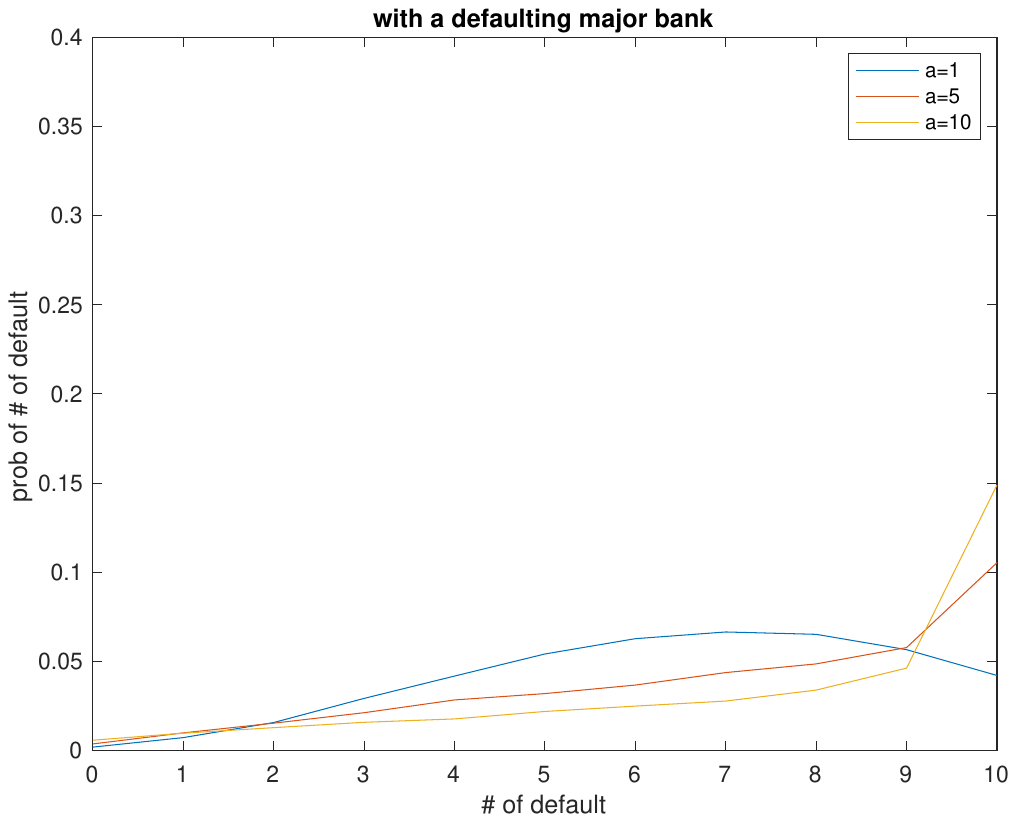}
    \end{minipage}
    }
    \caption{Loss distribution ($G=0.5,F=0.5$): (a) loss distribution for minor banks without major bank, (b) total loss distribution for minor banks with major bank, (c) loss distribution for minor banks conditional on the major bank not default, and (d) loss distribution for minor banks conditional on the major bank default.}
    \label{fig:LossDistribution_a}
\end{figure*}

Finally, we show the loss distributions of minor banks in the finite-population case in \cref{fig:LossDistribution_a}. Comparing panels (a) and (b) reveals that the tails become fatter with the mean-reversion parameter $a$, and even more so in the presence of a large bank. The distinction is clearer at the right tail which represents the scenario where all minor banks end up in default. 
These results are in line with \cref{RegressionEstimate_a_Finite:SystemicRisk} presented earlier. Panels (c) and (d) provide a clearer perspective on the role of the major bank. Firstly, the left tails in panels (a) and (c) exhibit a similar pattern, suggesting a comparable level of risk. However, the presence of a thicker right tail in (a) indicates a higher systemic risk in the absence of a stable major bank. On the other hand, panels (c) and (d) reveal distinct distributions. Specifically, when the major bank does not default, the distribution in panel (c) indicates a nearly negligible level of systemic risk. However, in the event of a major bank default as the mean reversion rate increases, the impact on systemic risk becomes more pronounced, as shown in panel (d). This emphasizes the significant role played by the mean reversion rate in influencing systemic risk during a major bank default scenario.

\end{document}